\documentclass[a4paper,11pt]{article}
\pdfoutput=1

\usepackage{jcappub} 
\usepackage{amsmath}
\usepackage{amssymb}
\usepackage{color}
\usepackage{graphicx}

\usepackage[utf8]{inputenc}

\graphicspath{{figures/}}

\begin{document}

\title{\boldmath Constraining early and interacting dark energy with gravitational wave standard sirens: the potential of the eLISA mission} 

\author[a]{Chiara Caprini}
\affiliation[a]{Institut de Physique Th\'eorique, CEA-Saclay, CNRS UMR 3681, Universit\'e Paris-Saclay, F-91191 Gif-sur-Yvette, France}
\emailAdd{chiara.caprini@cea.fr}

\author[a]{Nicola Tamanini}
\emailAdd{nicola.tamanini@cea.fr}

\date{\today}

\abstract{
We perform a forecast analysis of the capability of the eLISA space-based interferometer to constrain models of early and interacting dark energy using gravitational wave standard sirens. We employ simulated catalogues of standard sirens given by merging massive black hole binaries visible by eLISA, with an electromagnetic counterpart detectable by future telescopes. We consider three-arms mission designs with arm length of 1, 2 and 5 million km, 5 years of mission duration and the best-level low frequency noise as recently tested by the LISA Pathfinder. Standard sirens with eLISA give access to an intermediate range of redshift $1\lesssim z \lesssim 8$, and can therefore provide competitive constraints on models where the onset of the deviation from $\Lambda$CDM (i.e.~the epoch when early dark energy starts to be non-negligible, or when the interaction with dark matter begins) occurs relatively late, at $z\lesssim 6$. If instead early or interacting dark energy is relevant already in the pre-recombination era, current cosmological probes (especially the cosmic microwave background) are more efficient than eLISA in constraining these models, except possibly in the interacting dark energy model if the energy exchange is proportional to the energy density of dark energy.
}

\maketitle

\section{Introduction} 
\label{sec:introduction}

With the first direct detection of gravitational waves (GWs) by the LIGO/Virgo collaboration \cite{Abbott:2016blz,Abbott:2016nmj,TheLIGOScientific:2016pea} the era of GW astronomy has begun.
The information gathered from present and future GW observations will improve our understanding of the astrophysical objects emitting the GW signal, of the origin and evolution of the universe and its structure, and of the gravitational interaction.
Earth-based detectors, such as the advanced LIGO \cite{ligo} and Virgo \cite{virgo} interferometers, target the GW frequency window $10 - 1000$ Hz, while pulsar timing arrays (PTA), such as the ones united under the International Pulsar Timing Array (IPTA) collaboration \cite{2010CQGra..27h4013H}, probe much lower frequencies around $10^{-9} - 10^{-8}$ Hz.
In order to fill the gap in frequency between Earth-based interferometers and PTA, space-born GW observatories have been proposed, which will be able to reach high sensitivity in the frequency band $10^{-4} - 10^{-1}$ Hz.
Such range of frequencies in the GW landscape is supposed to be rich of astrophysical sources and it is yet completely unexplored.
In particular the strong GW signal emitted by merging massive black hole binaries (MBHBs) from $10^4$ to $10^7$ solar masses is expected to fall exactly within the frequency band targeted by space-born detectors.
Since such massive black holes are believed to reside at the centre of galaxies, observing the GW signal they emit will help to better understand the formation and evolution of galaxies and cosmic structures.

In 2013, the European Space Agency has approved a GW observatory in space as the L3 mission of its ``Cosmic Vision Program'' scheduled for launch around 2030-2034, for which the ``evolved LISA'' (eLISA) space-based interferometer is the main candidate \cite{elisaweb,Seoane:2013qna}.
eLISA is designed to probe the GW landscape around the mHz region where the signal produced by MBHBs is expected to be the loudest.
The final design of the mission, which is composed by three satellites orbiting around the Sun in an equilateral triangular formation, has not been decided yet, and some variables are still under considerations (see e.g. \cite{Klein:2015hvg}): the number of laser links between the satellites (four or six), corresponding to the number of active arms (two or three); the arm-length of the triangle (from one to five million km); and the duration of the mission (two or five years).
The eLISA low-frequency noise level (another of the variables previously considered) has been recently tested by the LISA Pathfinder mission \cite{pathfinderweb}, and according to the first results \cite{Armano:2016bkm} the expected noise is almost one hundred times better than the original requirement for the instrument.

In an earlier work, Ref.~\cite{Tamanini:2016zlh}, we have studied the capability of eLISA in probing the acceleration of the universe by means of MBHB mergers as {\it standard sirens}, i.e.~as sources of known distance \cite{schutz,Holz:2005df,Cutler:2009qv}. We have derived eLISA constraints on standard cosmological models: $\Lambda$CDM, dynamical dark energy, non-zero spatial curvature and so on. In the present paper, we specifically consider alternative scenarios to explain the acceleration of the universe: in particular, we study early and interacting dark energy models, see sections \ref{sec:early_DE} and \ref{sec:interacting_dark_energy}.
The principle of standard sirens is the following. The measured GW waveform depends directly on the luminosity distance of the source, and thus parameter estimation allows to infer the distance to the source for every GW event detected.
If subsequent electromagnetic (EM) observations are able to identify an EM counterpart, then one is able to obtain a measure of the source redshift and thus a point in the distance-redshift space.
Once a sufficient number of standard sirens is observed, the theoretically predicted distance-redshift relation can be compared against the data and constraints on the cosmological parameters can be statistically inferred. We assume spatial flatness throughout the paper, so that
\begin{equation}
	d_L(z) = c \left(1+z\right) \int_0^z \frac{1}{H(z')} dz' \,,
	\label{eq:dist_red_rel}
\end{equation}
where $c$ is the speed of light and $H(z)$ is the Hubble rate. 

The analysis presented here is meant to complete the one performed in \cite{Tamanini:2016zlh}. We concentrate on early (EDE) and interacting (IDE) dark energy, but use the same standard siren catalogues that have been obtained in \cite{Tamanini:2016zlh} starting from simulated rates of MBHB mergers detectable by different eLISA configurations, and considering realistic scenarios for the observation of the EM counterparts, based on the capabilities of future EM telescopes (LSST, SKA, ELT). Here we also consider the same three astrophysical models of MBHB formation and evolution appearing in \cite{Tamanini:2016zlh}, namely a light seeds model (popIII), a heavy seeds model with delay (Q3d) and a heavy seeds model without delay (Q3nod) (see also \cite{Klein:2015hvg} and references therein for more information). We present separate results for all these models. The number of standard sirens for each MBHB formation scenario has been selected under the hypothesis that the sky localisation of the event can be achieved using also the merger and ringdown phases of the signal. In this procedure the telescopes can be pointed only after the merger to look for a distinctive signature, therefore one implicitly assumes that there is a delay between the merger and the flare, or that the electromagnetic signal is persistent and peculiar enough that it can be confidently identified also minutes to hours after merger. This procedure was labelled the ``optimistic scenario'' in \cite{Tamanini:2016zlh}.
Moreover, the statistical methods employed to handle the simulated data coincide with the ones adopted in \cite{Tamanini:2016zlh}: in particular we perform a Fisher matrix analysis and obtain constraints and contour plots following the procedures exposed in \cite{Tamanini:2016zlh}. 

Differently from \cite{Tamanini:2016zlh}, in what follows we consider only three eLISA configurations, letting the arm-length to vary as one (A1), two (A2) and five (A5) million km, but fixing the number of laser links to six (L6), the mission duration to five years (M5) and the low-frequency noise to the LISA Pathfinder ``expected'' one (N2) (see \cite{Tamanini:2016zlh} and \cite{Klein:2015hvg} for details). The reasons for this choice are the following:
\begin{itemize}
	\item The aim of the present paper is not to carefully analyse all possible eLISA configurations to understand the science return of each of them (as it was in \cite{Tamanini:2016zlh}), but rather to investigate simple extensions of the $\Lambda$CDM model in order to understand the pros and cons of the eLISA mission in probing alternative cosmological models.
	\item According to the results of \cite{Tamanini:2016zlh}, four-link (two arms) configurations perform much worse than six-link (three arms) configurations in providing a sufficiently high number of MBHB standard sirens for cosmology. We therefore ignore four-link configurations since we expect that they will not be able to give meaningful constraints on the parameters of alternative cosmologies beyond $\Lambda$CDM.
	\item The number of detections, and thus the number of standard sirens, scales linearly with the mission duration: the longer the mission, the higher the number of datapoints. In analogy with the investigation of \cite{Tamanini:2016zlh} we thus only focus on a mission of five years\footnote{A method to estimate how the cosmological constraints change as the mission duration changes has been outlined in \cite{Tamanini:2016zlh}.}.
	\item Finally we only consider the ``expected'' low-frequency noise from LISA Pathfinder, called N2 \cite{Tamanini:2016zlh,Klein:2015hvg}. According to the first results of the mission \cite{Armano:2016bkm} this requirement has been met at frequencies $f>1$ mHz, while at lower frequencies the situation is still open: however one can optimistically forecast that the N2 noise level, if not a better one, will be finally achieved over the whole frequency spectrum.
\end{itemize}
The configuration with two million km arms, denoted N2A2M5L6, will be taken as our reference design for eLISA upon which the majority of subsequent results are based. When we need to pick a specific MBHB formation model, we choose the popIII scenario which is the one providing an intermediate number of standard sirens. 

In what follows we first investigate EDE in section \ref{sec:early_DE} and then analyse two different models of IDE in section \ref{sec:interacting_dark_energy}.
We have chosen EDE and IDE as alternative cosmological models because they are simple one-parameter extensions of $\Lambda$CDM and they allow to better expose the advantages of eLISA in probing the expansion of the Universe at high redshift. In the following, we set the fiducial values of the parameters to $\Omega_m^0=0.3$,~$w_0=-1$,~$h=0.67$,~ $\Omega_{de}^e=0$, $\epsilon_1=0$, $\epsilon_2=0$. Section \ref{sec:discussion_and_conclusion} contains discussions and conclusions.



\section{Early dark energy}
\label{sec:early_DE}

In early dark energy models, first proposed in \cite{Wetterich:2004pv}, the dark energy component evolves with redshift in such a way that it gives a non-negligible contribution also at early times, contrary to $\Lambda$CDM or other dynamical dark energy models where typically dark energy plays no role for redshift higher than about one. Early dark energy models can be probed by measuring the distance-redshift relation Eq.~\eqref{eq:dist_red_rel}. Ignoring the contribution of relativistic components and assuming spatial flatness, one can write
\begin{equation}
	\frac{H^2(z)}{H_0^2} = \frac{\Omega_m^0 (z+1)^3}{1 - \Omega_{de}(z)} \,,
	\label{eq:H}
\end{equation}
where $\Omega_m^0$ and $H_0$ are the present matter fraction and Hubble parameter and $\Omega_{de}(z)$ is the relative energy density of DE evolving in time, or equivalently over the redshift $z$: $\Omega_{de}(z)=\rho_{de}(z)/\rho_{tot}(z)$. 

The most widely used parameterization of $\Omega_{de}$ has been proposed in \cite{Doran:2006kp}:
\begin{equation}
 	\Omega_{de}(z) = \frac{\Omega_{de}^0 - \Omega_{de}^e \left[ 1- (z+1)^{3 w_0} \right]}{\Omega_{de}^0 + \Omega_{m}^0 (z+1)^{-3 w_0}} + \Omega_{de}^e \left[ 1 - (z+1)^{3 w_0} \right] \,,
 	\label{eq:early_DE}
\end{equation} 
where $w_0$ and $\Omega_{de}^0$ are the present values of respectively the equation of state (EoS) and relative energy density of dark energy, the latter being related to $\Omega_m^0$ through $\Omega_{de}^0 = 1 - \Omega_m^0$.
The parameter $\Omega_{de}^e$ characterizes the amount of dark energy present at early times $z \gg 1$, which remains constant also in the very early universe, before recombination. 

\begin{figure}
\begin{center}
	\includegraphics[width=.7\textwidth]{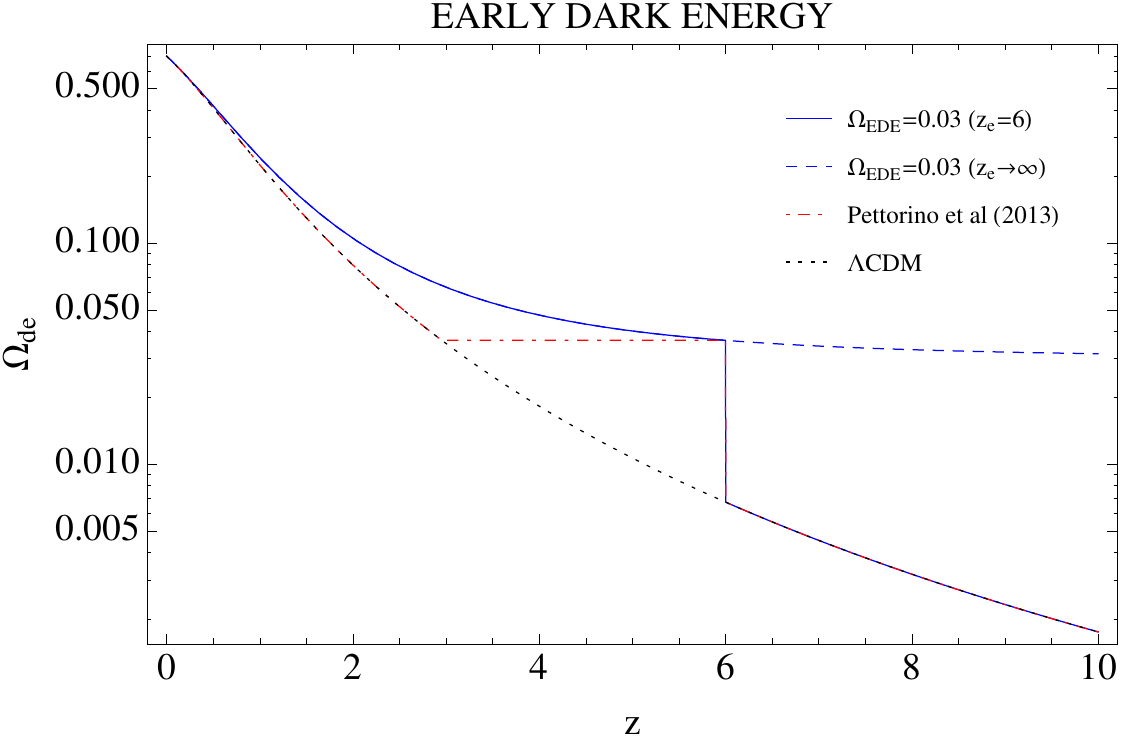}
\end{center}
\caption{Evolution of $\Omega_{de}(z)$ in early dark energy models with $\Omega_{de}^e = 0.03$ and $z_e = 6$ (solid blue line) or $z_e \rightarrow\infty$ (dashed blue line). The EDE model considered by Pettorino et al \cite{Pettorino:2013ia}, with the cut of EDE set to $z=6$, is also shown for comparison (dotted-dashed red line). The dotted black line represents $\Lambda$CDM.}
\label{fig:EDE_omega_z_plot}
\end{figure}

Here we analyse a parametrisation modified with respect to the one in Eq.~\eqref{eq:early_DE}, which allows us to investigate how the constraints change if $\Omega_{de}$ is non-negligible only for a limited amount of time in the past, instead of contributing beyond recombination as in the above model \eqref{eq:early_DE}. We will see that for eLISA the two models are equivalent as soon as the redshift at which early dark energy starts to contribute is sufficiently high: higher than about 6, as we will demonstrate. Therefore, even though the parametrisation in our analysis is different, we derive also eLISA constraints on the model \eqref{eq:early_DE}.

The scenario in which EDE is non-negligible only for a limited amount of time in the past
was first proposed in Ref.~\cite{Pettorino:2013ia}, which analyses how CMB constraints are affected by a variation of the epoch at which early dark energy starts to contribute. The model we consider here is somewhat different from those presented in~\cite{Pettorino:2013ia}, and shares some similarity with what done in Ref.~\cite{Aubourg:2014yra} in the case of Baryon Acoustic Oscillations (BAO). We let the universe become $\Lambda$CDM at redshift $z>z_e$, such that:
\begin{equation}
	\Omega_{de}(z) =
	\begin{cases}
		\frac{\Omega_{de}^0 - \Omega_{de}^e \left[ 1- (z+1)^{3 w_0} \right]}{\Omega_{de}^0 + \Omega_{m}^0 (z+1)^{-3 w_0}} + \Omega_{de}^e \left[ 1 - (z+1)^{3 w_0} \right] & \text{if } z < z_e \,, \\
		\frac{\Omega_{de}^0}{\Omega_{de}^0 + \Omega_m^0 (z+1)^3} & \text{if } z \geq z_e \,,
	\end{cases}
	\label{eq:not_so_early_DE}
\end{equation}
where $z_e$ determines the redshift up to which dark energy causes deviation from the usual $\Lambda$CDM expansion history, while for $z > z_e$ the standard $\Lambda$CDM evolution is recovered; see Fig.~\ref{fig:EDE_omega_z_plot}. This corresponds to model EDE3 of~\cite{Pettorino:2013ia} except that the universe does not go back to  $\Lambda$CDM at late time $a>a_c$: in EDE3 early dark energy is present only in the time interval between $z_c<z<z_e$, with $z_e$ a parameter and $z_c$ fixed by continuity; cf.~Fig.~\ref{fig:EDE_omega_z_plot}. Ref.~\cite{Aubourg:2014yra} instead considers a model in which the sound horizon is kept fixed at the fiducial $\Lambda$CDM value, meaning that early dark energy is negligible in the pre-recombination era and approaches the evolution of Eq.~\eqref{eq:early_DE} later in the matter era: this would correspond to our model \eqref{eq:not_so_early_DE} with sufficiently high $z_e$. 

The reason why we have chosen to consider model \eqref{eq:not_so_early_DE} is the following. 
As pointed out in \cite{Pettorino:2013ia}, CMB observations are mainly sensitive to deviations from $\Lambda$CDM at very high redshift: the CMB provides therefore very good constraints on $\Omega_{de}^e$ for the model \eqref{eq:early_DE}, as demonstrated also by the Planck analysis \cite{Ade:2015rim}. Ref.~\cite{Ade:2015rim} furthermore analyses the model EDE3 of \cite{Pettorino:2013ia}, where early dark energy is relevant only in the time interval between $z_c<z<z_e$: in this case, the constraint on $\Omega_{de}^e$ seriously degrades as $z_e$ decreases, because the CMB is less effective in constraining the late-time evolution of the universe (c.f. Fig.~11 of~\cite{Ade:2015rim}). Even though EDE3 is not exactly the same model as Eq.~\eqref{eq:not_so_early_DE}, one expects the CMB constraints obtained for EDE3 to equally apply to our parameterisation if $z_e$ is sufficiently small, say $z_e \lesssim 10$, precisely because CMB observations are mainly sensitive to deviations from $\Lambda$CDM only at very high redshift (and not at $z<z_e$). On the other hand, eLISA will be able to probe the redshift range $0 < z \lesssim 8$, because the redshift distribution of standard sirens extends in this interval: c.f. Fig.~\ref{fig:SS_z_distrib} of Appendix~\ref{sec:redshift_distribution_of_standard_sirens}, and the analysis of \cite{Tamanini:2016zlh}. We therefore expect any deviation from $\Lambda$CDM in the cosmic evolution happening at $z\lesssim 8$ to be best constrained by eLISA.
Hence the EDE model \eqref{eq:not_so_early_DE}, where the energy density of DE gives a non negligible contribution up to today, should be well tested by the eLISA mission. If instead the cosmic expansion history is not distinguishable from $\Lambda$CDM in the range $0 < z \lesssim 8$, as can happen in EDE3, no constraint can be put by eLISA on any parameter beyond $\Lambda$CDM (as $\Omega_{de}^e$). This is our main motivation for considering the parametrisation \eqref{eq:not_so_early_DE} as opposed to EDE3: it can be well constrained by eLISA, the constraints can be compared with those obtained using the CMB both if $z_e\rightarrow \infty$ and if $z_e \lesssim 10$, but at the same time they can also be compared with late-time constraints such as those, for example, given by BAO~\cite{Aubourg:2014yra} and 21-cm \cite{Archidiacono:2014msa}.  

In the following analysis we choose different values of $z_e$, namely $z_e = 1, 2, 3, 4, 6$ and $z_e\gg 6$: as we will see, this latter is effectively equivalent to $z_e\rightarrow \infty$, i.e.~to model \eqref{eq:early_DE} because eLISA will not be sensitive to transitions occurring after a redshift of about six. We investigate five cosmological models based on Eq.~\eqref{eq:not_so_early_DE}:
\begin{enumerate}
	\item A four-parameter model where every independent parameter is free: $\Omega_m^0$, $h$, $w_0$, $\Omega_{de}^e$;
	\item Three three-parameter models where one parameter among $\Omega_m^0$, $h$ and $w_0$ is fixed to its fiducial value and the others are free, together with $\Omega_{de}^e$;
	\item Three two-parameter models where one couple of parameters among $\Omega_m^0, w_0, h, \Omega_{de}^e$ is fixed to its fiducial values and the other couple is free;
	\item A one-parameter model where all $\Lambda$CDM parameters $\Omega_m^0$, $h$ and $w_0$ are fixed and only $\Omega_{de}^e$ is free.
\end{enumerate}
Note that since we are studying extensions to $\Lambda$CDM, $\Omega_{de}^e$ is always considered as a free parameter: fixing $\Omega_{de}^e$ to zero would reduce to the analysis already performed in \cite{Tamanini:2016zlh} where the resulting constraints on $\Omega_m^0$, $h$ and $w_0$ can be found.
On the other hand, $z_e$ is not taken as a free parameter but as a variable of the model.
Whenever $z_e\lesssim 10$, CMB observations cannot really constrain the model since they are sensitive only at high redshift (c.f. Fig.~11 of~\cite{Ade:2015rim}). Fixing the $\Lambda$CDM parameters to their fiducial values when $z_e\lesssim 10$ can therefore be considered equivalent to imposing a CMB prior.   

\begin{table}
\begin{center}
\begin{tabular}{|c|c|c|c||c|c|c|c|}
	\hline
	\multicolumn{8}{|c|}{EDE} \\
	\hline
	\multicolumn{4}{|c||}{$z_e = 2$} & \multicolumn{4}{|c|}{$z_e = 6$} \\
	\hline
	 $\Delta\Omega_m^0$ & $\Delta h$ & $\Delta w_0$ & $\Delta \Omega_{de}^e$ & $\Delta\Omega_m^0$ & $\Delta h$ & $\Delta w_0$ & $\Delta \Omega_{de}^e$ \\
	\hline
	 0.163 & 0.229 & 2.29 & 0.877 & 6.10 & 0.654 & 5.53 & 19.2 \\
	 0.280 & 0.450 & 3.80 & 0.983 & 1.08 & 0.255 & 1.87 & 3.12 \\
 	 0.0656 & 0.0728 & 0.797 & 0.399 & 1.48 & 0.160 & 1.52 & 4.60 \\
 	\hline
	\text{} & 0.0815 & 0.751 & 0.223 & \text{} & 0.0815 & 0.716 & 0.130 \\
 	\text{} & 0.0649 & 0.587 & 0.299 & \text{} & 0.144 & 0.877 & 0.228 \\
 	\text{} & 0.0457 & 0.450 & 0.146 & \text{} & 0.0342 & 0.336 & 0.0770 \\
 	\hline
 	0.0583 & \text{} & 0.193 & 0.313 & 0.826 & \text{} & 0.181 & 2.65 \\
    0.0427 & \text{} & 0.302 & 0.342 & 0.578 & \text{} & 0.291 & 1.90 \\
    0.0392 & \text{} & 0.121 & 0.192 & 0.349 & \text{} & 0.0928 & 1.08 \\
	\hline
 	0.0583 & 0.0190 & \text{} & 0.261 & 0.713 & 0.0175 & \text{} & 2.29 \\
 	0.0455 & 0.0377 & \text{} & 0.263 & 0.480 & 0.0421 & \text{} & 1.57 \\
 	0.0378 & 0.0102 & \text{} & 0.155 & 0.338 & 0.00932 & \text{} & 1.03 \\
 	\hline
 	0.0563 & \text{} & \text{} & 0.186 & 0.499 & \text{} & \text{} & 1.52 \\
 	0.0404 & \text{} & \text{} & 0.146 & 0.330 & \text{} & \text{} & 1.00 \\
 	0.0376 & \text{} & \text{} & 0.127 & 0.290 & \text{} & \text{} & 0.877 \\
	\hline
 	\text{} & 0.0188 & \text{} & 0.148 & \text{} & 0.0145 & \text{} & 0.102 \\
 	\text{} & 0.0370 & \text{} & 0.246 & \text{} & 0.0228 & \text{} & 0.129 \\
 	\text{} & 0.0102 & \text{} & 0.0806 & \text{} & 0.00876 & \text{} & 0.0631 \\
	\hline
 	\text{} & \text{} & 0.180 & 0.173 & \text{} & \text{} & 0.126 & 0.107 \\
 	\text{} & \text{} & 0.283 & 0.241 & \text{} & \text{} & 0.140 & 0.0988 \\
 	\text{} & \text{} & 0.117 & 0.109 & \text{} & \text{} & 0.0841 & 0.0719 \\
	\hline
 	\text{} & \text{} & \text{} & 0.0355 & \text{} & \text{} & \text{} & 0.0322 \\
 	\text{} & \text{} & \text{} & 0.0331 & \text{} & \text{} & \text{} & 0.0280 \\
 	\text{} & \text{} & \text{} & 0.0253 & \text{} & \text{} & \text{} & 0.0225 \\
	\hline
\end{tabular}
\end{center}
\caption{Standard 1$\sigma$ errors on early DE for N2A2M5L6. In the left table early DE is present only up to $z = 2$, while in the right table it is present only up to $z = 6$. In each row of the table, the top sub-row shows the errors for light seeds (popIII), the central sub-row for heavy seeds with delays (Q3d) and the bottom sub-row for heavy seeds without delays (Q3nod). Empty entries mean that the corresponding parameter has been fixed to its fiducial value (exact prior).}
\label{tab:err_EDE}
\end{table}

The forecast constraints that one can obtain with the eLISA configuration N2A2M5L6 are summarized in Table~\ref{tab:err_EDE} for $z_e = 2$ and $z_e = 6$. 
We stress that, since eLISA will probe only the redshift range up to $z \simeq 8$, any model with $z_e\gtrsim 8$ will have the same constraints. In practice, as we will see, our analysis shows that already after $z\simeq 6$ the constraints on $\Omega_{de}^e$ stabilize: for this reason $z_e \gtrsim 6$ or $z_e \rightarrow \infty$ are equivalent from the point of view of the constraints.

\begin{figure}
\begin{center}
	\includegraphics[width=\textwidth]{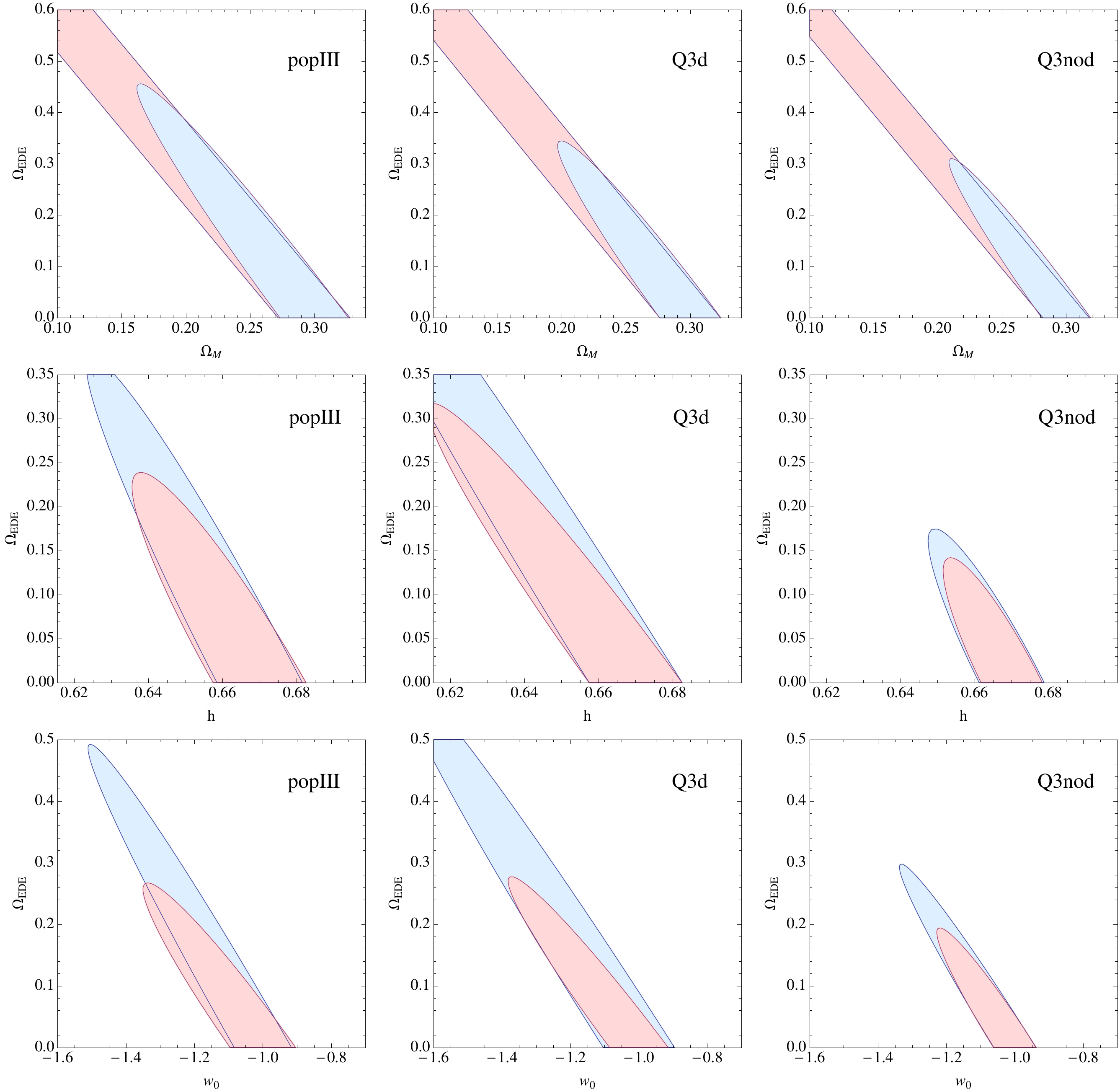}
\end{center}
\caption{EDE: 2$\sigma$ contours for $z_e = 2$ (blue) and $z_e = 6$ (red) with N2A2M5L6 for the three MBHB formation scenarios in the two-parameter cosmological models where $\Omega_{de}^e$ is a free parameter together with $\Omega_m^0$, $h$ and $w_0$, respectively.}
\label{fig:early_DE_ellipses}
\end{figure}

Notice in Table~\ref{tab:err_EDE} the difference between the cases $z_e = 2$ and $z_e = 6$ when all four parameters are free to vary (first row): the case $z_e = 2$ is poorly, but at least slightly constrained, while the case $z_e = 6$ is not at all constrained (errors bigger than 100\% of the parameter fiducial values).
This is due to the fact that the parameters $\Omega_m^0$ and $\Omega_{de}^e$ are degenerate (as shown by Eqs.~\eqref{eq:H} and \eqref{eq:early_DE}), combined with the fact that the redshift distribution of standard sirens extends up to $z\simeq 8$ and there are in general few events at low redshift (this depends somewhat on the MBHB formation model: see Appendix~\ref{sec:redshift_distribution_of_standard_sirens} for the redshift distribution of standard sirens).
For $z>z_e$, the standard sirens data are effectively constraining a three parameter model because $\Omega_{de}^e$ is zero and the expansion is the same as $\Lambda$CDM. Fixing $z_e=2$ allows therefore for a better measurement of the three $\Lambda$CDM parameters from the majority of the standard sirens data at high redshift, and in turn also of $\Omega_{de}^e$ due to the degeneracy with $\Omega_m^0$.
On the other hand, when $z_e = 6$ the majority of the standard sirens data are effectively constraining the full four-parameter cosmological model, thus the accuracy with which the parameters can be determined is worse than in the $z_e = 2$ case. 

This effect can be understood also by comparing for example the 5th to 7th rows of Table~\ref{tab:err_EDE} (two-parameter models).
Fixing both $h$ and $w_0$ to their fiducial values (5th row of Table~\ref{tab:err_EDE}) improves the constraints with respect to the four parameter model but does not break the degeneracy among $\Omega_m^0$ and $\Omega_{de}^e$: for this reason, the case $z_e = 6$ still has larger errors than the case $z_e = 2$.
On the other hand, when $\Omega_m^0$ is fixed together with another parameter (rows six and seven of Table~\ref{tab:err_EDE}), the case $z_e = 6$ becomes better constrained than the case $z_e = 2$: the degeneracy has been broken by fixing $\Omega_m^0$ and more standard sirens are available if $z_e = 6$ to help constrain the EDE model. The same behaviour is observed in rows two to four of Table~\ref{tab:err_EDE} (three-parameter models), although it is less evident and can depend on the MBHB formation scenario.   

In Fig.~\ref{fig:early_DE_ellipses} we show 2$\sigma$ contour plots for the two-parameter models $(\Omega_{de}^e,~\Omega_m^0)$, $(\Omega_{de}^e,~h)$ and $(\Omega_{de}^e,~w_0)$. The figure represents all three MBHB models for the configuration N2A2M5L6, and both cases $z_e = 2$ and $z_e = 6$. The contour plots in the $(\Omega_{de}^e, \Omega_{m}^0)$ plane clearly show the degeneracy between these two parameters and how is it improved by choosing $z_e = 2$ instead of $z_e = 6$. On the other hand, when $\Omega_{m}^0$ is fixed to its fiducial value (which at these redshift can be considered equivalent to setting a CMB prior), the degeneracy with $\Omega_{de}^e$ is broken and the constraints improve; however, there remains some level of degeneracy among $\Omega_{de}^e$ and respectively $h$ and $w_0$, as can be appreciated from the second and third rows of figure~\ref{fig:early_DE_ellipses}. 

\begin{figure}
\begin{center}
	\includegraphics[width=\textwidth]{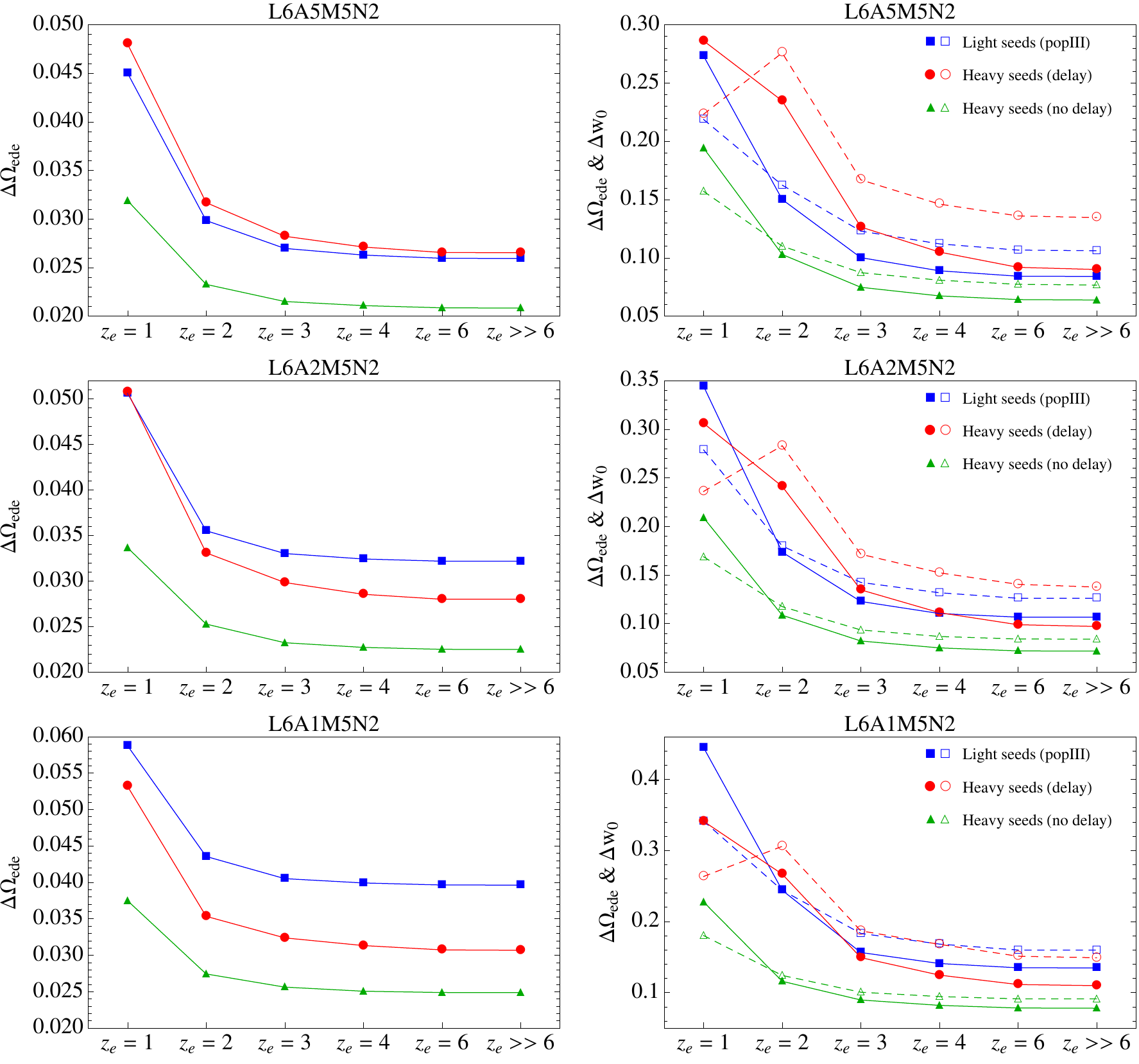}
\end{center}
\caption{1$\sigma$ errors in the one-parameter cosmological model with only $\Omega_{de}^e$ (left panels) and in the two-parameter cosmological models with both $\Omega_{de}^e$ and $w_0$ (right panel) for three 6-link eLISA configurations. In the right panels empty and filled markers denote the uncertainties on $w_0$ and $\Omega_{de}^e$, respectively.}
\label{fig:early_DE}
\end{figure}

Fig.~\ref{fig:early_DE} shows how the accuracy in determining the parameters $\Omega_{de}^e$ and $w_0$ changes with different values of $z_e$.
We consider the three eLISA configurations with 6-links, noise N2 and varying arm-length, and the three MBHB formation models. The left column shows the one-parameter model with $\Omega_{de}^e$ only and the right column the two-parameter model with both $\Omega_{de}^e$ and $w_0$. 
Since $\Omega_{m}^0$ is fixed, the errors always decrease with increasing $z_e$ because of the higher number of standard sirens available for the measurement\footnote{Note that the error on $w_0$ in the model heavy seeds with delay (Q3d) actually increases going from $z_e = 1$ to $z_e = 2$. This is due to the fact that for Q3d there is a very low number of standard sirens with $z < 1$ (see Appendix~\ref{sec:redshift_distribution_of_standard_sirens}). Thus, when $z_e=1$, almost all the data are effectively constraining a one-parameter model where $\Omega_{de}^e=0$ and only $w_0$ is left free to vary, providing in this manner a good constraint on $w_0$. However, for $z_e \geq 2$ the number of standard sirens constraining the full two-parameter model is sufficiently high so that both $\Delta\Omega_{de}^e$ and $\Delta w_0$ decrease as $z_e$ increases. This suggests that there exists a redshift between 1 and 2 where the error on $w_0$ is the highest. The fact that this effect does not appear in the other BH models is due to the higher number of standard sirens at low redshift: it shows that in those cases the redshift at which $\Delta w_0$ is largest happens to be below $z = 1$ .} (cf.~Appendix~\ref{sec:redshift_distribution_of_standard_sirens}). However they stabilise around $z_e=6$ and do not change appreciably if $z_e\gg 6$: the constraining power of eLISA cannot improve further due to the reduced number of standard sirens after $z \simeq 6$ (c.f. redshift distributions in Appendix~\ref{sec:redshift_distribution_of_standard_sirens}). 

\subsection*{Comparison with present constraints}

The errors on $\Omega_{de}^e$ for $z_e \geq 6$ when all other parameters are held fixed are more than one order of magnitude worse than the ones presently available from CMB observations \cite{Hojjati:2013oya}, in particular if compared with the latest Planck results \cite{Ade:2015rim} which provide a 2$\sigma$ uncertainty on $\Omega_{de}^e$ of 0.0036. In the same regime, i.e.~when early dark energy is relevant back to the pre-recombination era and the sound horizon is therefore rescaled, BAO measurements do not improve on CMB constraints since they are plagued, as we are, by the strong degeneracy between $\Omega_m^0$ and $\Omega_{de}^e$. Forecasts for 21-cm probes instead give a constraint better than 10\% on $\Omega_{de}^e$, according to \cite{Archidiacono:2014msa}.  

In summary, if $z_e \gg 10$, present CMB experiments already perform way better than what eLISA will be able to provide. 
On the other hand, CMB observations are unable to give significant constraints whenever $z_e \lesssim 10$ \cite{Pettorino:2013ia,Ade:2015rim}, while the constraints on $\Omega_{de}^e$ by eLISA outlined in Table~\ref{tab:err_EDE} become competitive when $z_e$ is sufficiently low. This highlights the main strength of eLISA in testing alternative cosmological models: if deviations from $\Lambda$CDM occur only in the range $z \lesssim 6$, eLISA has higher constraining ability than CMB probes and can therefore be considered as complementary to them. If instead $z_e\gg 6$, eLISA cannot compete with CMB probes. Concerning BAO measurements, they provide stronger constraints than eLISA only in the case when the sound horizon is held fixed at the fiducial $\Lambda$CDM value (i.e.~early dark energy becomes relevant at some point in the matter-dominated era): Ref.~\cite{Aubourg:2014yra} finds $\Delta \Omega_{de}^e= 0.031$ at 2$\sigma$. If this is not the case, present BAO observations perform worse than future eLISA.
Finally SNIa observations are not expected to significantly improve constraints on early dark energy because they rely on measurements at low redshift.


\section{Interacting dark energy} 
\label{sec:interacting_dark_energy}

Interacting dark energy (IDE) models have been first proposed to help alleviate the coincidence problem \cite{Wetterich:1994bg,Amendola:1999er}, and have been widely studied in the literature because they seem to be favoured by present cosmological data, especially when redshift space distortions (RSD) are included in the datasets used to establish the constraints (c.f. discussions in Sec. \ref{sec:IDE1} and \ref{sec:IDE2}). In IDE one introduces a coupling between DE and DM that at the background level modifies the conservation equations as
\begin{align}
	\dot\rho_{dm} + 3 H \rho_{dm} &= Q \,, \label{eq:IDE_cons_dm} \\
	\dot\rho_{de} + 3 H (1+w_0) \rho_{de} &= -Q \,, \label{eq:IDE_cons_de}
\end{align}
where an over-dot stands for differentiation with respect to cosmic time and $Q$ defines the amount of energy exchanged between the two dark fluids.
In what follows we neglect the baryonic and radiation contributions and consider two terms for the energy exchange, that we denote respectively IDE1 and IDE2:
\begin{equation}
	Q = \epsilon_1 H \rho_{dm}~~ \text{(IDE1)} \quad\quad\text{and}\quad\quad Q = \epsilon_2 H \rho_{de}~~ \text{(IDE2)} \,.
	\label{eq:IDE_Qs}
\end{equation}
These are the simplest phenomenological models of IDE and have been extensively investigated in the literature (see e.g. \cite{Wang:2016lxa} for a recent review).
The purpose of the present analysis is to test the ability of eLISA in constraining a possible interaction in the dark sector, not to distinguish between different interacting models: we therefore do not consider more elaborated IDE models.

Note that standard sirens, similarly to SNIa, only probe the expansion of the universe at the background level and thus in our analysis there is no need to specify the fully covariant form of the interactions defined by Eq.~\eqref{eq:IDE_Qs}.
We therefore do not need to worry about possible covariantization issues \cite{Li:2014eha,Faraoni:2014vra,Tamanini:2015iia,Skordis:2015yra} or instabilities at the perturbation level \cite{Valiviita:2008iv,He:2008si,Jackson:2009mz}. Consequently, we leave the parameters $\epsilon_1$ and $\epsilon_2$ free to take both positive and negative values, and their sign does not have to be connected to the value of $w_0$: these restrictions are necessary for analyses that need to perturb the dark fluids because they take into account  observational probes of cosmological perturbations (CMB, BAO, RSD...) \cite{He:2008si,Valiviita:2008iv,Gavela:2009cy,Marcondes:2016reb}. It is clear that any realistic model of IDE needs to be stable  under perturbations, but here we prefer to show the plain result of our analysis without setting any stability prior on parameters. 
Since standard sirens only probe the background, their constraining power is not optimal compared to other combined probes, as we will see; however, we have the advantage of being general, in the sense that any IDE model which reduces to \eqref{eq:IDE_Qs} at the background level, independently of the full form of its interacting exchange four-vector, can be constrained by the results that follow.

\begin{figure}
\begin{center}
	\includegraphics[width=.7\textwidth]{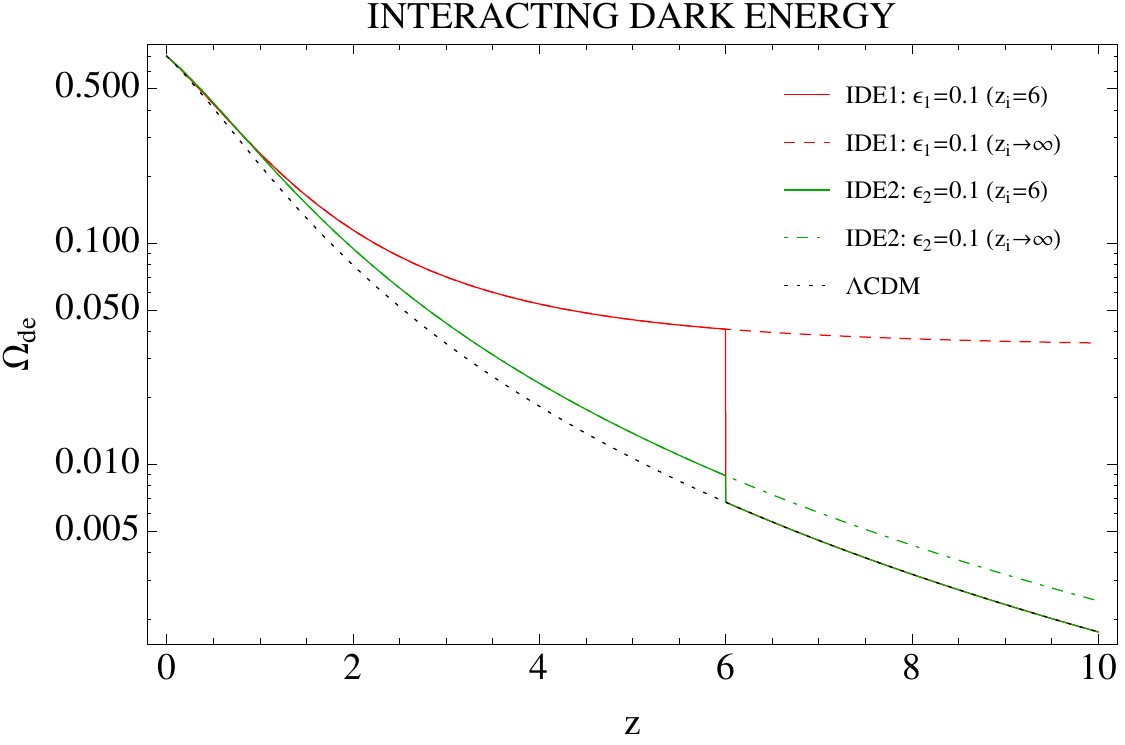}
\end{center}
\caption{Evolution of $\Omega_{de}(z)$ in interacting dark energy models. From top to bottom of the curves: the red line denotes IDE1 with $\epsilon_1 = 0.1$ and $z_i = 6$ (solid line) or $z_i \rightarrow\infty$ (dashed line). The green line denotes IDE2 with $\epsilon_2 = 0.1$ and $z_i = 6$ (solid line) or $z_i \rightarrow\infty$ (dotted-dashed line). The black, dotted line denotes $\Lambda$CDM.}
\label{fig:IDE_omega_z_plot}
\end{figure}

In both cases of Eq.~\eqref{eq:IDE_Qs}, Eqs.~\eqref{eq:IDE_cons_dm}--\eqref{eq:IDE_cons_de} can be solved analytically yielding
\begin{align}
	\rho_{dm} &= \rho_{dm}^0 (1+z)^{3- \epsilon_1} \,,\\
	\rho_{de} &= \rho_{de}^0 (1+z)^{3(1+w_0)} + \frac{\epsilon_1}{\epsilon_1+3 w_0} \rho_{dm}^0 \left[ (1+z)^{3(1+w_0)} - (1+z)^{3- \epsilon_1} \right] \,,
\end{align}
for IDE1 and
\begin{align}
	\rho_{dm} &= \rho_{dm}^0 (1+z)^3 + \rho_{de}^0 (1+z)^3 \left[ \frac{\epsilon_2}{\epsilon_2 + 3 w_0} \left( 1 - (1+z)^{3 w_0 + \epsilon_2} \right) \right] \,,\\
	\rho_{de} &= \rho_{de}^0 (1+z)^{3(1+w_0) + \epsilon_2} \,,
\end{align}
for IDE2, where $\rho_{dm}^0$ and $\rho_{de}^0$ are the values of the energy densities at $z=0$. For negative values of $\epsilon_1$ the energy flows from dark matter to dark energy, and for positive values of $\epsilon_2$ the energy flows from dark energy to dark matter. 
Inserting the above equations into the Friedmann equation one obtains
\begin{equation}
	\frac{H^2}{H_0^2} = \Omega_{m}^0 (1+z)^{3- \epsilon_1} + \Omega_{de}^0 (1+z)^{3 (1+w_0)} + \Omega_{m}^0 \frac{\epsilon_1}{\epsilon_1 + 3 w_0} \left[ (1+z)^{3 (1+w_0)} - (1+z)^{3- \epsilon_1} \right] \,,
	\label{eq:H_IDE1}
\end{equation}
for IDE1 and
\begin{equation}
	\frac{H^2}{H_0^2} = \Omega_{m}^0 (1+z)^3 + \Omega_{de}^0 (1+z)^3 \left[ \frac{\epsilon_2}{\epsilon_2 + 3 w_0} \left(1 - (1+z)^{3 w_0 + \epsilon_2} \right) \right] + \Omega_{de}^0 (1+z)^{3(1+w_0) + \epsilon_2} \,,
	\label{eq:H_IDE2}
\end{equation}
for IDE2, which have to be inserted in the distance-redshift relation \eqref{eq:dist_red_rel} in order to fit the standard sirens data. Note that in both IDE models the interaction is between dark energy and dark matter only, without baryons. Therefore in Eq.~\eqref{eq:H_IDE1} $\Omega_{m}^0$ should in reality be the parameter $\Omega_{dm}^0=\rho_{dm}^0/\rho_{\rm tot}^0$. Here we neglect this fact and write Eq.~\eqref{eq:H_IDE1} in terms of the total $\Omega_{m}^0$: this introduces an error, but since standard sirens only probe the background evolution and not the growth of structure we expect this error to be small. 

As for EDE, we consider IDE models where the interaction between dark energy and dark matter is negligible for redshift higher than some reference redshift $z_i$, see e.g.~\cite{Cai:2009ht,Salvatelli:2014zta}.
For such models the Hubble rate is given by Eq.~\eqref{eq:H_IDE1} (or Eq.~\eqref{eq:H_IDE2}) only up to $z_i$, while for higher redshift it is set to its $\Lambda$CDM behaviour.
In other words the evolution of $\Omega_{de}(z) = \rho_{de}(z)/\rho_{\rm tot}(z)$ (and consequently of $\Omega_{dm}$) is modified only up to $z_i$, and goes to $\Lambda$CDM for higher redshifts; see Fig.~\ref{fig:IDE_omega_z_plot}.
Similarly to EDE, a late time interaction in the dark sector cannot be efficiently constrained by CMB data (see e.g.~\cite{Salvatelli:2014zta}). This is again the main motivation for considering IDE models with a given $z_i$: in this case standard sirens data, probing the expansion in the redshift range $0< z\lesssim 8$, can reveal useful to strengthen the constraints derived from CMB. For example, Ref.~\cite{Salvatelli:2014zta} analyses the IDE2 model fixing $w_0=-1$ and letting the interaction parameter $\epsilon_2$ take different values in different redshift bins: for a single bin (corresponding to our model) with $z_i=0.9$, they find that a null interaction is excluded at 99\% confidence level. This result is obtained combining Planck data with RSD, and shows that an IDE model where the interaction switches on at late times is a possible solution to the
tension that arises in $\Lambda$CDM between CMB and RSD data. 

In the following analysis we consider again different values of $z_i$, namely $z_i = 1, 2, 3, 4, 6$ and $z_i\gg 6$: with the latter effectively equivalent to $z_i\rightarrow \infty$, for the same reasons discussed in Sec.~\ref{sec:early_DE} for EDE. 
For both IDE models, we investigate the following scenarios:
\begin{enumerate}
	\item A four-parameter model where every parameter is free: $\Omega_m^0$, $h$, $w_0$ and $\epsilon_1$ (or $\epsilon_2$);
	\item Three three-parameter models where one parameter among $\Omega_m^0$, $h$ and $w_0$ is fixed to its fiducial value and the others are free, together with $\epsilon_1$ (or $\epsilon_2$);
	\item Three two-parameter models where one couple of parameters among $\Omega_m^0, w_0, h, \epsilon_1 (\text{or}~\epsilon_2)$ is fixed to its fiducial values and the other couple is free;
	\item A one-parameter model where all $\Lambda$CDM parameters $\Omega_m^0$, $h$ and $w_0$ are fixed and only $\epsilon_1$ (or $\epsilon_2$) is free.
\end{enumerate}
We stress again that since we are studying extensions to $\Lambda$CDM, $\epsilon_1$ and $\epsilon_2$ are always considered as free parameters: fixing $\epsilon_1$ (or $\epsilon_2$) to zero would reduce the analysis to the standard cosmology one performed in \cite{Tamanini:2016zlh}.

\subsection{IDE1}
\label{sec:IDE1}

\begin{table}
\begin{center}
\begin{tabular}{|c|c|c|c||c|c|c|c|}
	\hline
	\multicolumn{8}{|c|}{IDE1: $Q = \epsilon_1 H \rho_{dm}$} \\
	\hline
	\multicolumn{4}{|c||}{$z_i = 2$} & \multicolumn{4}{|c|}{$z_i = 6$} \\
	\hline
	 $\Delta\Omega_m^0$ & $\Delta h$ & $\Delta w_0$ & $\Delta \epsilon_1$ & $\Delta\Omega_m^0$ & $\Delta h$ & $\Delta w_0$ & $\Delta \epsilon_1$ \\
	\hline
	 0.148 & 0.168 & 1.13 & 1.04 & 0.695 & 0.251 & 3.25 & 2.52 \\
	 0.287 & 0.395 & 2.73 & 1.21 & 0.492 & 0.302 & 3.09 & 1.77 \\
 	 0.0660 & 0.0511 & 0.423 & 0.485 & 0.333 & 0.0760 & 1.32 & 1.13 \\
	\hline
	\text{} & 0.0840 & 0.649 & 0.274 & \text{} & 0.0811 & 0.663 & 0.139 \\
 	\text{} & 0.0681 & 0.514 & 0.351 & \text{} & 0.196 & 1.40 & 0.291 \\
 	\text{} & 0.0393 & 0.373 & 0.177 & \text{} & 0.0339 & 0.277 & 0.0838 \\
	\hline
 	0.0740 & \text{} & 0.147 & 0.530 & 0.258 & \text{} & 0.526 & 0.866 \\
 	0.0518 & \text{} & 0.142 & 0.452 & 0.283 & \text{} & 0.641 & 0.880 \\
 	0.0503 & \text{} & 0.0954 & 0.344 & 0.152 & \text{} & 0.313 & 0.535 \\
	\hline
 	0.0841 & 0.0195 & \text{} & 0.578 & 0.145 & 0.0367 & \text{} & 0.490 \\
 	0.0622 & 0.0204 & \text{} & 0.458 & 0.201 & 0.0621 & \text{} & 0.551 \\
 	0.0561 & 0.0118 & \text{} & 0.377 & 0.0758 & 0.0180 & \text{} & 0.282 \\
	\hline
 	0.0431 & \text{} & \text{} & 0.440 & 0.0297 & \text{} & \text{} & 0.204 \\
 	0.0428 & \text{} & \text{} & 0.407 & 0.0298 & \text{} & \text{} & 0.178 \\
 	0.0296 & \text{} & \text{} & 0.295 & 0.0200 & \text{} & \text{} & 0.137 \\
	\hline
 	\text{} & 0.00889 & \text{} & 0.217 & \text{} & 0.00736 & \text{} & 0.117 \\
 	\text{} & 0.0157 & \text{} & 0.283 & \text{} & 0.00931 & \text{} & 0.103 \\
 	\text{} & 0.00571 & \text{} & 0.136 & \text{} & 0.00479 & \text{} & 0.0786 \\
	\hline
 	\text{} & \text{} & 0.0794 & 0.228 & \text{} & \text{} & 0.0608 & 0.116 \\
 	\text{} & \text{} & 0.118 & 0.262 & \text{} & \text{} & 0.0672 & 0.0979 \\
 	\text{} & \text{} & 0.0553 & 0.155 & \text{} & \text{} & 0.0415 & 0.0795 \\
	\hline
 	\text{} & \text{} & \text{} & 0.104 & \text{} & \text{} & \text{} & 0.0722 \\
 	\text{} & \text{} & \text{} & 0.0886 & \text{} & \text{} & \text{} & 0.0564 \\
 	\text{} & \text{} & \text{} & 0.0738 & \text{} & \text{} & \text{} & 0.0513 \\
	\hline
\end{tabular}
\end{center}
\caption{Standard 1$\sigma$ errors on the parameters of IDE1 for the eLISA configuration N2A2M5L6. In the left table the interaction is present only up to $z = 2$, while in the right table it is present only up to $z = 6$. In each row of the table, the top sub-row shows the errors for light seeds (popIII), the central sub-row for heavy seeds with delays (Q3d) and the bottom sub-row for heavy seeds without delays (Q3nod). Blank entries mean that the corresponding parameter has been fixed to its fiducial value (exact prior).}
\label{tab:err_IDE1}
\end{table}
\begin{figure}
\begin{center}
	\includegraphics[width=\textwidth]{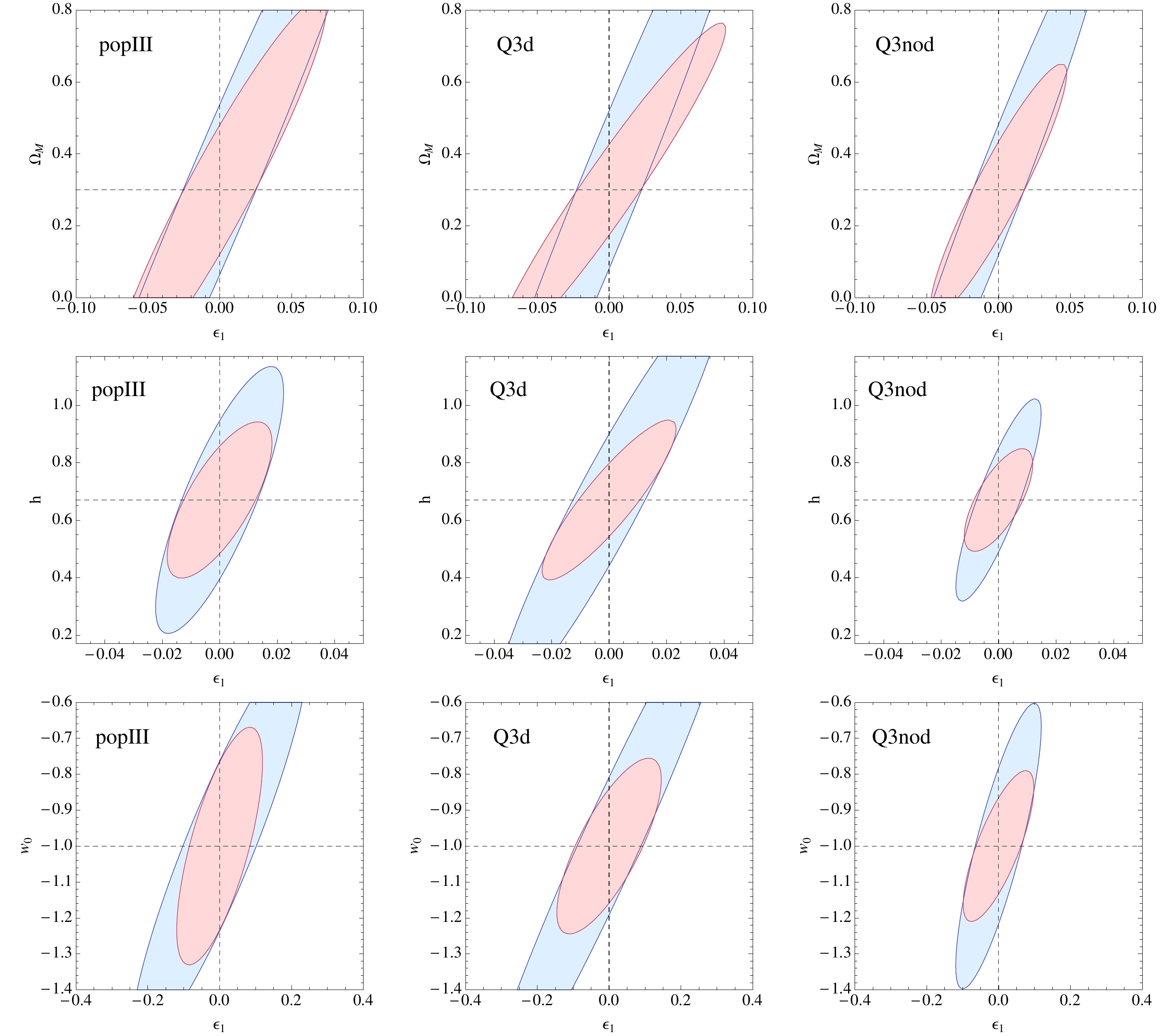}
\end{center}
\caption{IDE1: 2$\sigma$ contours for $z_i = 2$ (blue) and $z_i = 6$ (red) with N2A2M5L6 for three MBHB scenarios in the two-parameter cosmological models where $\epsilon_1$ is a free parameter together with $\Omega_{m}^0$, $h$ and $w_0$, respectively.}
\label{fig:ellipses_IDE1}
\end{figure}

Let us start from IDE1. Standard 1$\sigma$ errors for the eLISA configuration N2A2M5L6 are shown in Table~\ref{tab:err_IDE1}.
The situation is quite similar to the EDE case of Sec.~\ref{sec:early_DE}: because of the strong degeneracy between $\Omega_{m}^0$ and $\epsilon_1$, the errors in the four-parameter model increase if IDE is present up to higher redshift, as shown in the first row of Table~\ref{tab:err_IDE1}. The $z_i = 2$ errors are smaller than the $z_i = 6$ ones because in the first case the data are effectively constraining a three-parameter model (with parameters $\Omega_m^0$, $h$ and $w_0$) for $z > 2$, where the majority of standard sirens is (see Appendix~\ref{sec:redshift_distribution_of_standard_sirens}), partly bypassing degeneracies between $\epsilon_1$ and the other parameters. Moreover, Table~\ref{tab:err_IDE1} shows that whenever $\Omega_{m}^0$ is fixed to its fiducial value, the errors do decrease going from $z_i = 2$ to $z_i = 6$, as expected: this is clear in rows six and seven of Table~\ref{tab:err_IDE1}, while for the three parameter model (second row) it is less evident, as was the case in the EDE model. On the other hand, if either $h$ or $w_0$ (but not $\Omega_{m}^0$) is set to its fiducial value in the three-parameter models, the degeneracy among the parameters remains, as can be appreciated in the 3rd and 4th rows of Table~\ref{tab:err_IDE1}, where the errors are still bigger in the $z_i = 6$ case. In Appendix~\ref{sec:degeneracies} we present the marginalised contour plots for the three-parameter models when $z_i \gtrsim 6$ (see Fig.~\ref{fig:degeneracies_1}) and discuss further the degeneracies of these models. 

Fig.~\ref{fig:ellipses_IDE1} shows contour plots in the two-parameter models $(\epsilon_1,\Omega_{m}^0)$, $(\epsilon_1,h)$ and $(\epsilon_1,w_0)$, for the three MBHB formation scenarios, configuration N2A2M5L6, and both $z_e = 2$ and $z_e = 6$. A relevant difference with the EDE case appears in the two-parameter model $(\epsilon_1,\Omega_{m}^0)$: even though $\Omega_{m}^0$ is free to vary together with $\epsilon_1$, the errors are smaller in the $z_i = 6$ case, contrary to what happened for EDE. This can be appreciated both comparing the ellipses in the first row of Fig.~\ref{fig:ellipses_IDE1} with those in the first row of Fig.~\ref{fig:early_DE_ellipses}, and from the values of the errors in row five of Table~\ref{tab:err_IDE1}. The degeneracy among $\Omega_{m}^0$ and $\epsilon_1$ is still evident but less serious than in the EDE case.

\begin{figure}
\begin{center}
	\includegraphics[width=\textwidth]{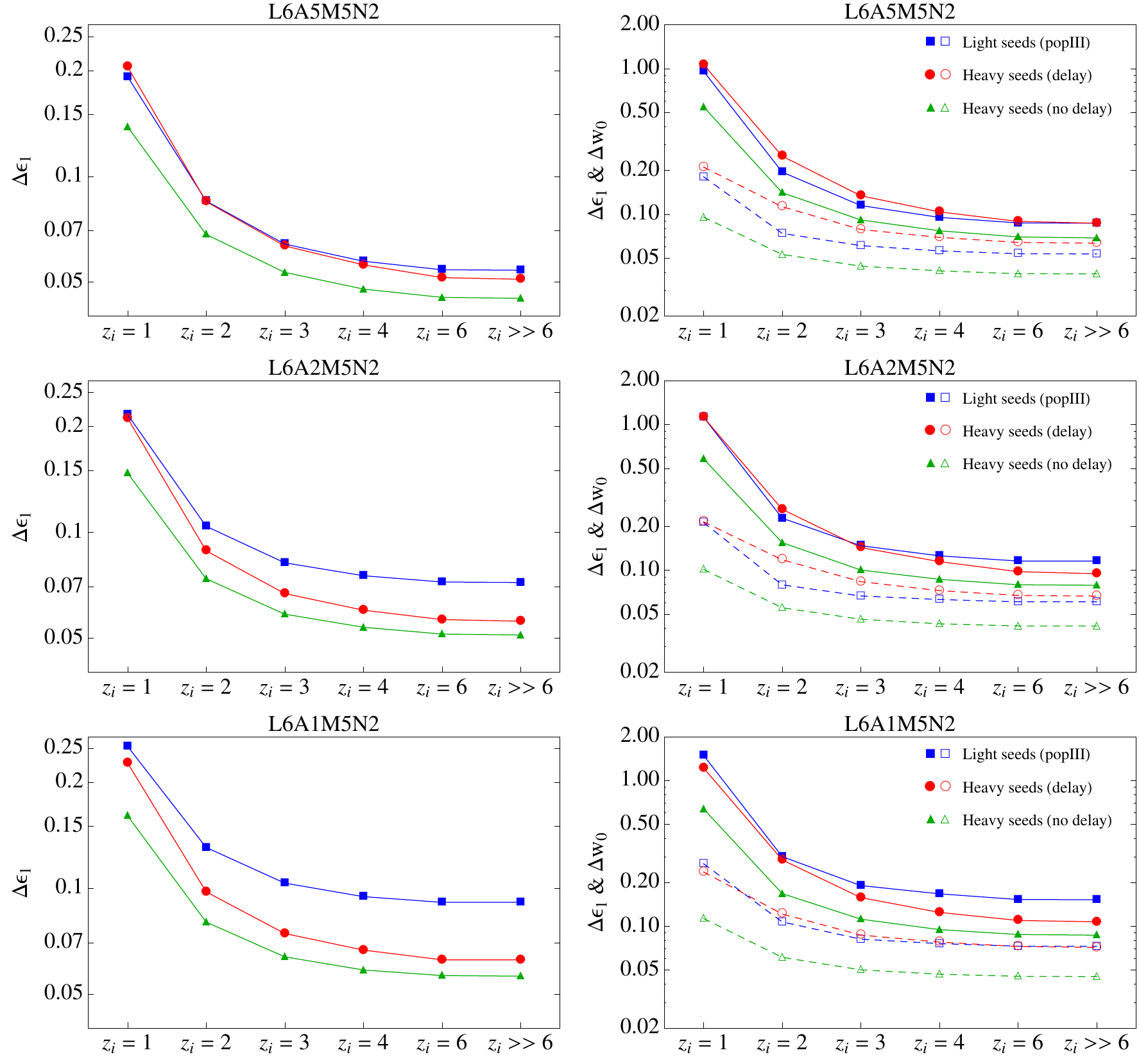}
\end{center}
\caption{IDE1: 1$\sigma$ errors on $\epsilon_1$ and $w_0$ in the one parameter cosmological model with only $\epsilon_1$ (left panels) and in the two parameter cosmological models with both $\epsilon_1$ and $w_0$ (right panel) for three different 6-link eLISA configurations. In the right panels empty and filled markers denote the uncertainties on $w_0$ and $\epsilon_1$, respectively.}
\label{fig:IDE_1}
\end{figure}

In Fig.~\ref{fig:IDE_1} we compare models with different $z_i$ by means of 1$\sigma$ errors on $\epsilon_1$ and $w_0$.
As for the EDE case, we consider the three eLISA configurations with 6-links, noise N2 and varying arm-length, and the three MBHB formation scenarios. The left column represents the one-parameter model with $\epsilon_1$ and the right column the two-parameter model with both $\epsilon_1$ and $w_0$. As expected when fixing $\Omega_{m}^0$, the constraints on both parameters improve as $z_i$ grows\footnote{Differently to what observed for EDE, in this case the errors on $w_0$ decrease going from $z_i = 1$ to $z_i = 2$ also in the heavy seeds with delay scenario (Q3d). However, the gain in $\Delta w_0$ going from $z_i = 1$ to $z_i = 2$ is less pronounced than in the Q3nod and PopIII models. As for EDE, this is again due to the lower number of standard sirens present in the Q3d model at low redshift (see Appendix~\ref{sec:redshift_distribution_of_standard_sirens}), but here the effect is not strong enough to make $\Delta w_0$ at $z_i = 1$ smaller than at $z_i = 2$, as it happens for EDE (cf.~Sec.~\ref{sec:early_DE}).}. They stabilise around $z_i=6$ and do not change appreciably if $z_i\gg 6$: the constraining power of eLISA cannot improve further due to the reduced number of standard sirens after $z \simeq 6$. 

\subsubsection*{Comparison with present constraints}

The present status of constraints on the IDE1 model is that the cosmological data show a tendency to prefer a small positive interaction $\epsilon_1>0$ with $w<-1$, although the significance is often low and depends on the combination of datasets chosen for the analysis. We give some examples of latest analyses, which only consider the case where the interaction is present also far in the past, $z_i\rightarrow \infty$. Ref.~\cite{Costa:2016tpb} analyses the IDE1 model with $\lambda_1=\epsilon_1/3$ and finds $\lambda_1=0.0006628^{+ 0.000241}_{-0.000592} $ and $w_0=-1.069^{+ 0.0268}_{-0.0152}$ at 1$\sigma$, when all probes are combined (Planck+SNIa+BAO+$H_0$+RSD). Also Ref.~\cite{Nunes:2016dlj} finds a mild preference for a positive $\epsilon_1$ and $w<-1$, but only when considering exclusively $H_0$ probes supplemented by cosmic chronometers techniques. Refs. \cite{Sola:2016ecz,Li:2015vla} analyse models of interacting vacuum energy, setting then $w_0=-1$. The first work, Ref.~\cite{Sola:2016ecz}, finds that the SNIa+BAO+$H(z)$+LSS+Planck data favour a mild dynamical vacuum evolution\footnote{Here LSS means the measurement of $\sigma_8$.}, while Ref.~\cite{Li:2015vla} uses again Planck+SNIa+BAO+$H_0$ data to find a strong constraint on $\beta=\epsilon_1/3$ with no evidence of a positive interaction: $ \beta = -0.00045\pm 0.00069 $. Therefore, constraints on IDE1 still depend widely on the chosen combination of datasets and on the data analysis technique. However, we remark that in general the claimed sensitivity on $\epsilon_1$ of present cosmological probes is higher by at least two orders of magnitude than the forecast for eLISA we give in the present work with all other parameters fixed. Furthermore, this sensitivity may be largely improved by the time eLISA will fly, in particular with the help of the Euclid survey. The current analyses only consider models for which the interaction is present also far in the past, and we cannot therefore compare with them the case in which eLISA provides its best constraints. However, given the big discrepancy in the precision of the measurement for $z_i\rightarrow \infty$, it is likely that for IDE1 eLISA will only serve as an independent but not so sensitive mean to test the model, and possibly break degeneracies with other cosmological probes when one considers the scenario in which the interaction between dark energy and dark matter is negligible for redshift higher than some reference redshift $z_i$.


\subsection{IDE2}
\label{sec:IDE2}

\begin{table}
\begin{center}
\begin{tabular}{|c|c|c|c||c|c|c|c|}
	\hline
	\multicolumn{8}{|c|}{IDE2: $Q = \epsilon_2 H \rho_{de}$} \\
	\hline
	\multicolumn{4}{|c||}{$z_i = 2$} & \multicolumn{4}{|c|}{$z_i = 6$} \\
	\hline
	 $\Delta\Omega_m^0$ & $\Delta h$ & $\Delta w_0$ & $\Delta \epsilon_2$ & $\Delta\Omega_m^0$ & $\Delta h$ & $\Delta w_0$ & $\Delta \epsilon_2$ \\
	\hline
	 0.168 & 0.199 & 1.56 & 1.23 & 2610 & 0.0977 & 3730 & 11200 \\
	 0.308 & 0.448 & 3.22 & 1.49 & 2520 & 0.165 & 3590 & 10800 \\
 	 0.0692 & 0.0622 & 0.534 & 0.545 & 1540 & 0.0384 & 2200 & 6600 \\
 	\hline
 	\text{} & 0.0819 & 0.662 & 0.295 & \text{} & 0.0803 & 0.686 & 0.178 \\
 	\text{} & 0.0668 & 0.522 & 0.398 & \text{} & 0.148 & 1.02 & 0.308 \\
 	\text{} & 0.0421 & 0.398 & 0.192 & \text{} & 0.0338 & 0.307 & 0.105 \\
	\hline
 	0.0682 & \text{} & 0.128 & 0.495 & 5.93 & \text{} & 8.45 & 25.4 \\
 	0.0481 & \text{} & 0.157 & 0.486 & 0.844 & \text{} & 1.20 & 3.67 \\
 	0.0455 & \text{} & 0.0851 & 0.322 & 1.36 & \text{} & 1.94 & 5.87 \\
	\hline
 	0.0751 & 0.0156 & \text{} & 0.524 & 0.487 & 0.0843 & \text{} & 2.16 \\
 	0.0558 & 0.0229 & \text{} & 0.441 & 0.567 & 0.126 & \text{} & 2.32 \\
 	0.0502 & 0.00937 & \text{} & 0.340 & 0.217 & 0.0343 & \text{} & 0.993 \\
 	\hline
 	0.0569 & \text{} & \text{} & 0.466 & 0.0541 & \text{} & \text{} & 0.366 \\
 	0.0455 & \text{} & \text{} & 0.374 & 0.0632 & \text{} & \text{} & 0.396 \\
 	0.0386 & \text{} & \text{} & 0.315 & 0.0374 & \text{} & \text{} & 0.248 \\
	\hline
 	\text{} & 0.0112 & \text{} & 0.217 & \text{} & 0.00916 & \text{} & 0.144 \\
 	\text{} & 0.0208 & \text{} & 0.332 & \text{} & 0.0128 & \text{} & 0.148 \\
 	\text{} & 0.00688 & \text{} & 0.132 & \text{} & 0.00580 & \text{} & 0.0932 \\
	\hline
 	\text{} & \text{} & 0.101 & 0.239 & \text{} & \text{} & 0.0772 & 0.146 \\
 	\text{} & \text{} & 0.154 & 0.296 & \text{} & \text{} & 0.0903 & 0.130 \\
 	\text{} & \text{} & 0.0697 & 0.157 & \text{} & \text{} & 0.0534 & 0.0994 \\
	\hline
 	\text{} & \text{} & \text{} & 0.0866 & \text{} & \text{} & \text{} & 0.0703 \\
 	\text{} & \text{} & \text{} & 0.0765 & \text{} & \text{} & \text{} & 0.0577 \\
 	\text{} & \text{} & \text{} & 0.0609 & \text{} & \text{} & \text{} & 0.0499 \\
	\hline
\end{tabular}
\end{center}
\caption{Standard 1$\sigma$ errors on IDE2 parameters for N2A2M5L6. In the left table the interaction is present only up to $z = 2$, while in the right table it is present only up to $z = 6$. In each row of the table, the top sub-row shows the errors for light seeds (popIII), the central sub-row for heavy seeds with delays (Q3d) and the bottom sub-row for heavy seeds without delays (Q3nod). Blank entries mean that the corresponding parameter has been fixed to its fiducial value (exact prior).}
\label{tab:err_IDE2}
\end{table}
\begin{figure}
\begin{center}
	\includegraphics[width=\textwidth]{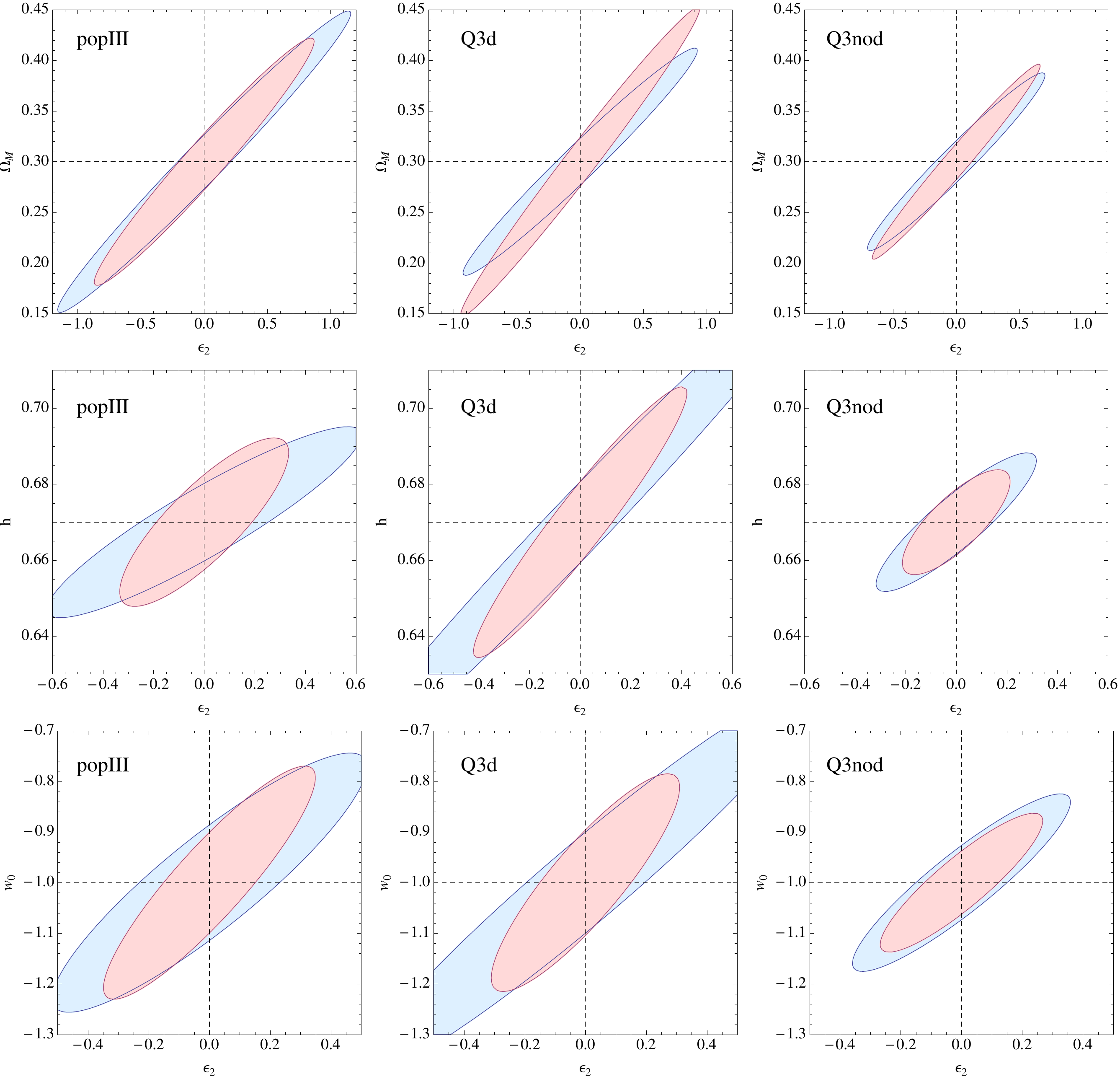}
\end{center}
\caption{IDE2: 2$\sigma$ contours for $z_i = 2$ (blue) and $z_i = 6$ (red) with N2A2M5L6 for three MBHB models in the two-parameter cosmological models where $\epsilon_2$ is a free parameter together with $\Omega_{m}^0$, $h$ and $w_0$, respectively.}
\label{fig:ellipses_IDE2}
\end{figure}

Standard 1$\sigma$ errors for the IDE2 model are shown in Table~\ref{tab:err_IDE2} where, as usual, we have chosen the eLISA configuration N2A2M5L6. In this model we also expect a degeneracy between $\Omega_{m}^0$ and $\epsilon_2$ due to the energy exchange between dark energy and dark matter and vice-versa. Table~\ref{tab:err_IDE2} shows that the degeneracy is more severe than in the IDE1 case: while the errors on all parameters are comparable with those of IDE1 when $z_i=2$, they are much larger when $z_i=6$. This is especially patent in the four-parameter model, but one can see from the table that in general IDE2 with any number of free parameter is more loosely constrained than IDE1 if $z_i=6$. Note in particular the two-parameter model $(\epsilon_2,\Omega_{m}^0)$, fifth row of Table~\ref{tab:err_IDE2}: in the heavy seed with delay scenario (the one with the lowest number of standard sirens), the errors are higher when $z_i=6$ than when $z_i=2$, meaning that the degeneracy is not broken contrary to what happened in the IDE1 case. This can be appreciated also from the first line of Fig.~\ref{fig:ellipses_IDE2}, where we show the 2$\sigma$ contour plots for all the two-parameter models in the three different MBHB formation scenarios. Degeneracies in the IDE2 model are further discussed in Appendix~\ref{sec:degeneracies}, and in Figs.~\ref{fig:degeneracies_1} and \ref{fig:degeneracies_2} of the Appendix we report marginalised contour plots for three-parameter cosmological models. 

\begin{figure}
\begin{center}
	\includegraphics[width=\textwidth]{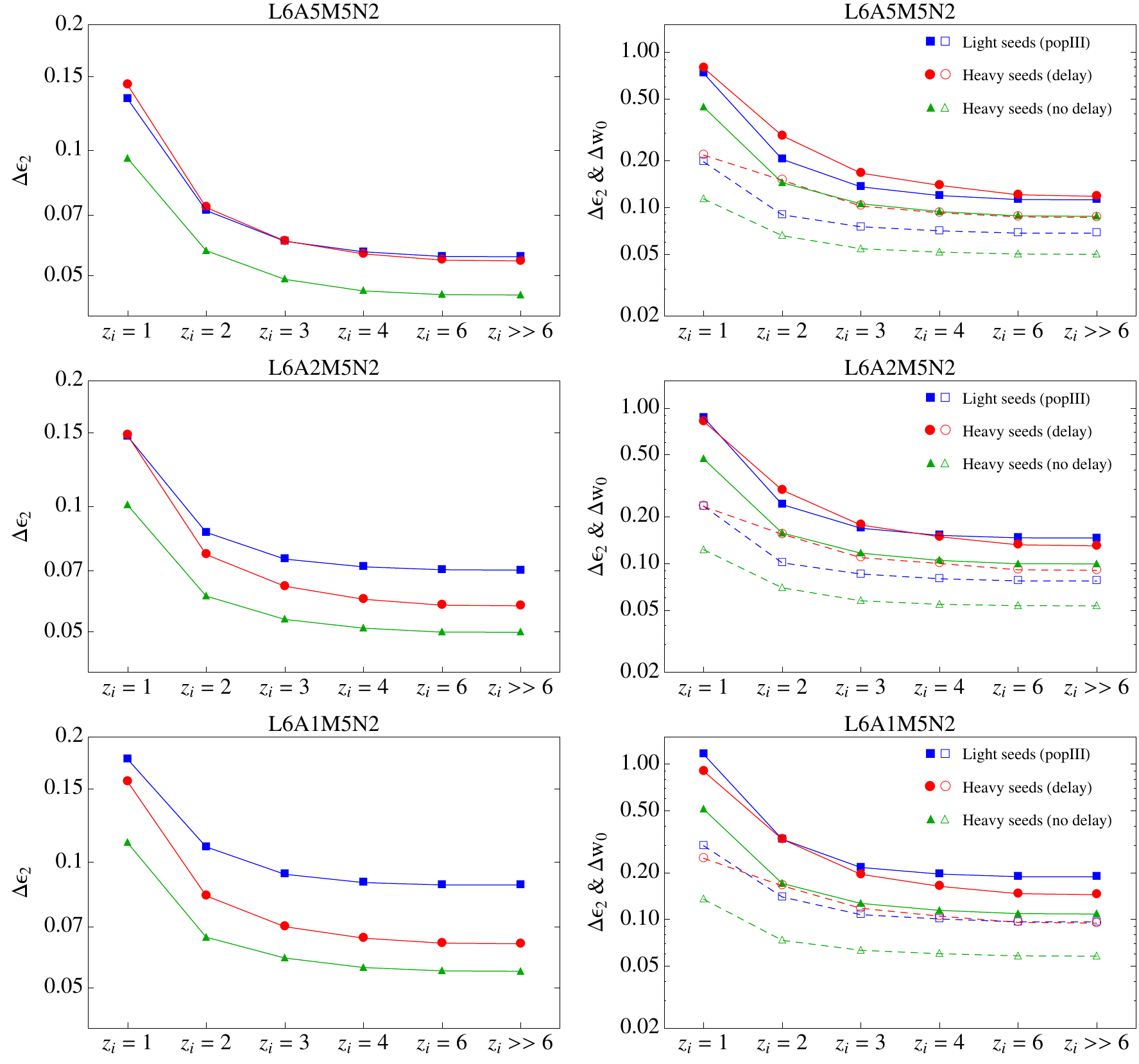}
\end{center}
\caption{IDE2: 1$\sigma$ constraints on $\epsilon_2$ and $w_0$ in the one parameter cosmological model with only $\epsilon_2$ (left panels) and in the two parameter cosmological models with both $\epsilon$ and $w_0$ (right panel) for three different 6-link eLISA configurations. In the right panels empty and filled markers denote the uncertainties on $w_0$ and $\epsilon_2$, respectively.}
\label{fig:IDE_2}
\end{figure}

Like for EDE and IDE1, a comparison between IDE2 models with different $z_i$ is given in Fig.~\ref{fig:IDE_2} in terms of errors on $\epsilon_2$ and $w_0$: the analysis of this figure is analogous to what discussed for IDE1. 

\subsubsection*{Comparison with present constraints}

The first evidence that IDE2 could be preferred by cosmological observations was presented in Ref.~\cite{Salvatelli:2013wra}. Subsequently Ref.~\cite{Salvatelli:2014zta} analysed the model of a late-time interaction between dark matter and vacuum energy, i.e.~IDE2 with $w_0=-1$ and $\epsilon_2$ varying in several redshift bins. They found moderate
evidence for a negative interaction ($\epsilon_2<0$) starting at $z_i=0.9$ from the combination of Planck, SNIa and RSD data, excluding the null interaction ($\Lambda$CDM) at 99\%. On the other hand, Ref.~\cite{Murgia:2016ccp} considers the case of an interaction extending back in the past, $z_i\rightarrow\infty$, exclude $w_0=-1$ to avoid instabilities and finds that a positive coupling with $w_0<-1$ is favoured by the combination of Planck, SNIa and BAO/RSD data: they get $\epsilon_2=0.159^{+0.146}_{-0.154}$ at 2$\sigma$. This result is qualitatively consistent with Ref.~\cite{Costa:2016tpb}, which also excludes a zero positive interaction and finds $\lambda_2=\epsilon_2/3=0.02047^{+0.00565}_{-0.00667}$ at 1$\sigma$ (using also $H_0$ measurements), and with Ref.~\cite{Feng:2016djj} where Planck, SNIa, BAO and $H_0$ data are used to get $\lambda_2=\epsilon_2/3=0.0782^{+0.0377}_{-0.0347}$ at 2$\sigma$. Ref.~\cite{Sola:2016ecz} also finds evidence for a non-zero coupling (although here $w_0$ is fixed to -1). Conversely, Ref.~\cite{Li:2015vla} again finds no evidence for a non-zero interaction, as was the case for IDE1: with Planck+SNIa+BAO+$H_0$ data, the authors constrain $\beta=\epsilon_1/3$ to $ \beta = -0.026^{+0.036}_{-0.053}$ when $z_i \rightarrow\infty$ and fixing $w_0 = -1$. It is important to point out that overall, the constraints and/or error bars on the IDE2 model from present observational datasets are less stringent than for the IDE1 model, because of degeneracies: for example, the same analysis applied to both models in Ref.~\cite{Li:2015vla} led to constraints on IDE2 that are weaker by two orders of magnitude. As we have stressed, degeneracies do degrade also the eLISA forecasts, but only by a factor of a few: we expect therefore that eLISA, in combination with other cosmological observables, will be able to improve the constraints on IDE2, in particular for those models assuming low values of $z_i$.  


\section{Discussion and conclusion} 
\label{sec:discussion_and_conclusion}

We have presented a forecast analysis of the capabilities of the eLISA mission to constrain two alternative models of dark energy, namely early and interacting dark energy.  These models have been widely studied in the literature, and previous analyses have shown the advantages of testing them using data at various redshifts, combining different observational probes such as CMB, BAO, SNIa, LSS, weak lensing, RSD. The motivation of this work resides in the fact that standard sirens with eLISA can provide access to an intermediate range of redshift $1\lesssim z \lesssim 8$, higher than what can be reached with SNIa and matter structure data. Furthermore, the measurement of the luminosity distance with standard sirens being given by gravitational waves, it provides an independent access to the distance-redshift relation, partly complementary to electromagnetic observations. 

In the present analysis we have used the same procedure developed in \cite{Tamanini:2016zlh}: we have started from simulations of the event rates of MBHB in three different models for the BH seeds, and we have used realistic scenarios for the occurrence and detection of the EM counterparts. Equipped with catalogues of standard sirens, we have selected those which are visible by different eLISA configurations setting the thresholds of SNR$>8$ and sky localisation better than 10 ${\rm deg}^2$. These have been achieved including both the inspiral and merger and ringdown phases of the GW event (the ``optimistic scenario'' in  \cite{Tamanini:2016zlh}). Since in Ref.~ \cite{Tamanini:2016zlh} we demonstrated that eLISA configurations with four links are not very powerful in probing the expansion of the universe, here we concentrated only on six-link configurations. We have also fixed the duration of the mission to five years (as done in \cite{Tamanini:2016zlh}) and the noise level to the Pathfinder expected one (N2), which is justified after the success of the LISA Pathfinder mission \cite{Armano:2016bkm}. We have therefore considered three eLISA configurations: N2A1M5L6, N2A2M5L6, N2A5M5L6. 

The main result of the present analysis is that standard sirens with eLISA can be competitive in constraining EDE and IDE models if the onset of the deviation from $\Lambda$CDM (i.e.~the epoch when EDE starts to be non-negligible, or when the interaction with DM begins) occurs relatively late, at $z \lesssim 6$. Models for which the deviation from $\Lambda$CDM starts far in the past, typically before recombination, are well constrained by current cosmological probes, in particualr by the CMB; the present constraints are way better than those that can be achieved with even the best configurations of eLISA (except perhaps for the IDE2 case, depending on the analysis one compares with: c.f. discussion at the end of Sec.~\ref{sec:IDE2}). On the other hand, if the deviation starts relatively late, the present observational constraints on both EDE and IDE models are highly degraded, and eLISA becomes therefore competitive in testing these scenarios. 

This happens because the redshift distribution of standard sirens peaks between $2\leq z \leq 4$, and very few standard sirens are available at redshift larger than six. The errors on the EDE and IDE parameters beyond $\Lambda$CDM (namely $w_0$, $\Omega_{de}^e$, $\epsilon_{1}$, $\epsilon_2$), when all the other cosmological parameters are held fixed, decrease with the increase of the redshift at which the deviation from $\Lambda$CDM starts. However, this occurs up to a redshift for the onset of the deviation of about six; for higher deviation redshift, the eLISA errors stabilize and do not change appreciably up to far in the radiation era. This behaviour of the errors follows the redshift distribution of the standard sirens (see appendix~\ref{sec:redshift_distribution_of_standard_sirens}) which are available for the measurement: after a redshift of about six, the number of standard sirens detected does not increase sufficiently any longer to provide a better measurement of the cosmological parameters. 

We have also demonstrated, however, that this behaviour can be affected by degeneracies among the parameters, in particular among $\Omega_m^0$ and $\Omega_{de}^e$ or $\epsilon_{1}$, $\epsilon_2$ (a degeneracy which is expected in both EDE and IDE models due to the way these parameters enter in the distance-redshift relation). If $\Omega_m^0$ is not set to its fiducial value, the errors on $\Omega_{de}^e$ or $\epsilon_{1}$, $\epsilon_2$ increase when the redshift of the onset of the deviation from $\Lambda$CDM increases. Once again this reflects the fact that the peak of the standard siren distribution resides in the interval  $2\leq z \leq 4$ (see appendix~\ref{sec:redshift_distribution_of_standard_sirens}): for low deviation redshift, the bulk of the MBHB standard sirens detected by eLISA effectively probes $\Lambda$CDM or dynamical dark energy (i.e.~with $w_0\neq -1$), and this in turns leads to a reduction of the errors on $\Omega_{de}^e$ or $\epsilon_{1}$, $\epsilon_2$. This happens especially for the models where either all four, or three parameters are free to vary. Therefore, without setting an exact prior on $\Omega_m^0$, eLISA can only constrain models where the redshift of the onset of the deviation from $\Lambda$CDM is sufficiently low. Note however, that whenever the deviation from $\Lambda$CDM starts well after recombination, from the point of view of the eLISA analysis one can confidently use the very precise CMB measurement of $\Omega_m^0$ as an exact prior. 

We can therefore conclude that eLISA with six-link configurations will serve as an independent mean to test alternative models for the acceleration of the universe such as EDE and IDE, and will be able to improve the present constraints, in particular for the EDE and IDE2 models, if one considers low values of the redshift at which the deviation from $\Lambda$CDM starts.

\acknowledgments

We thank the {\it Institut d'Astrophysique de Paris} and the institute {\it AstroParticule et Cosmologie} at {\it Universit\'{e} Paris Diderot} for hospitality.  
We also thank Enrico Barausse for useful comments on the draft.
NT acknowledge support from the Labex P2IO and the Enhanced Eurotalents Programme.


\appendix

\section{Redshift distribution of MBHB standard sirens} 
\label{sec:redshift_distribution_of_standard_sirens}

\begin{figure}
\begin{center}
	\includegraphics[width=\textwidth]{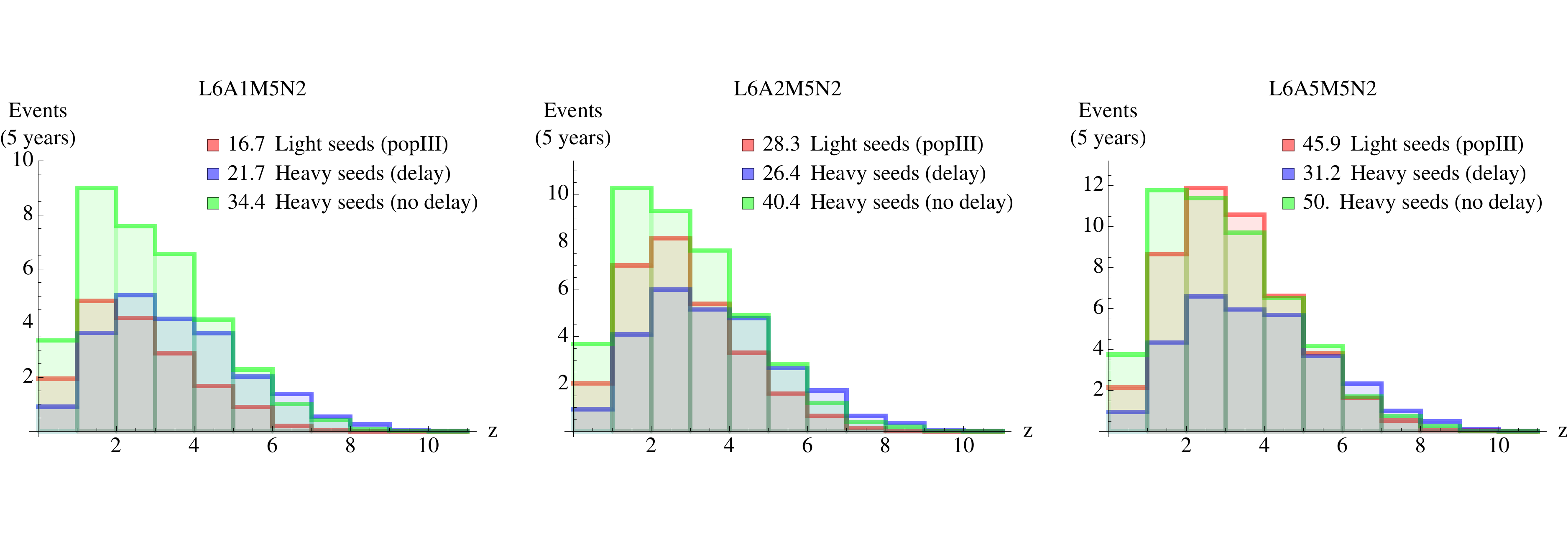}
\end{center}
\caption{Redshift distribution of MBHB standard sirens for the three different astrophysical models considered. Numbers in the legends denote the total amount of standard sirens detected over all redshift bins.}
\label{fig:SS_z_distrib}
\end{figure}

In this appendix we provide the distribution in redshift of the number of MBHB standard sirens obtained from the analysis performed in \cite{Tamanini:2016zlh}.
This serves as supporting material for some of the statements made in the main text and completes the results presented in \cite{Tamanini:2016zlh}.

In Fig.~\ref{fig:SS_z_distrib} we present the distribution in redshift of standard sirens for the three best eLISA configurations, and the three MBHB formation models considered.
We stress that only GW events for which the redshift can be measured from an EM counterpart are shown in Fig.~\ref{fig:SS_z_distrib}.
The total number of GW events detected by eLISA, with and without counterpart observation, can be found in \cite{Tamanini:2016zlh}.
From Fig.~\ref{fig:SS_z_distrib} one first notices that for all MBHB models the distribution is peaked for $2 \lesssim z \lesssim 3$, which thus represents the redshift range around which the bulk of MBHB standard sirens data are expected.
Note also that the distributions have tails reaching even $z=10$, meaning that MBHB standard sirens will be able to directly test the expansion at very high redshifts, though for datapoints with $z \gtrsim 6$ a large error bar is expected due to weak lensing effects and the difficulties in measuring the redshift from an optical counterpart.

In general standard sirens from Q3nod events are more abundant than Q3d ones, reflecting the fact that Q3nod and Q3d are respectively optimistic and pessimistic scenarios for heavy seed models.
This is especially true at very low redshifts ($z < 1$), where there are almost no events for Q3d.
Going from N2A1M5L6 to N2A5M5L6, for light seeds there is a relative gain with respect to heavy seed models in the total number of standard sirens, and a shift of the distribution peak towards higher redshift.
This is due to the fact that MBHB mergers are usually less massive in light seed models and more abundant if compared with heavy seeds.
The situation can be better understood looking at Table 10 of \cite{Tamanini:2016zlh}.
For the three best configurations, the total number of GW detections remains constant in both heavy seed models, while it greatly improves for light seeds. 
This happens because heavy seed mergers are more massive and thus all events can be detected already with N2A1M5L6.
On the other hand light seed mergers are lighter and thus N2A1M5L6 can only detect a small part of them, while N2A5M5L6 is able to observe a larger amount.
This in turn is reflected on the total number of standard sirens which, going from N2A1M5L6 to N2A5M5L6, increases in light seeds much more rapidly than in heavy seed models since more GW events are available.


\section{Degeneracies in three-parameter IDE models} 
\label{sec:degeneracies}

\begin{figure}
\begin{center}
	\includegraphics[width=\textwidth]{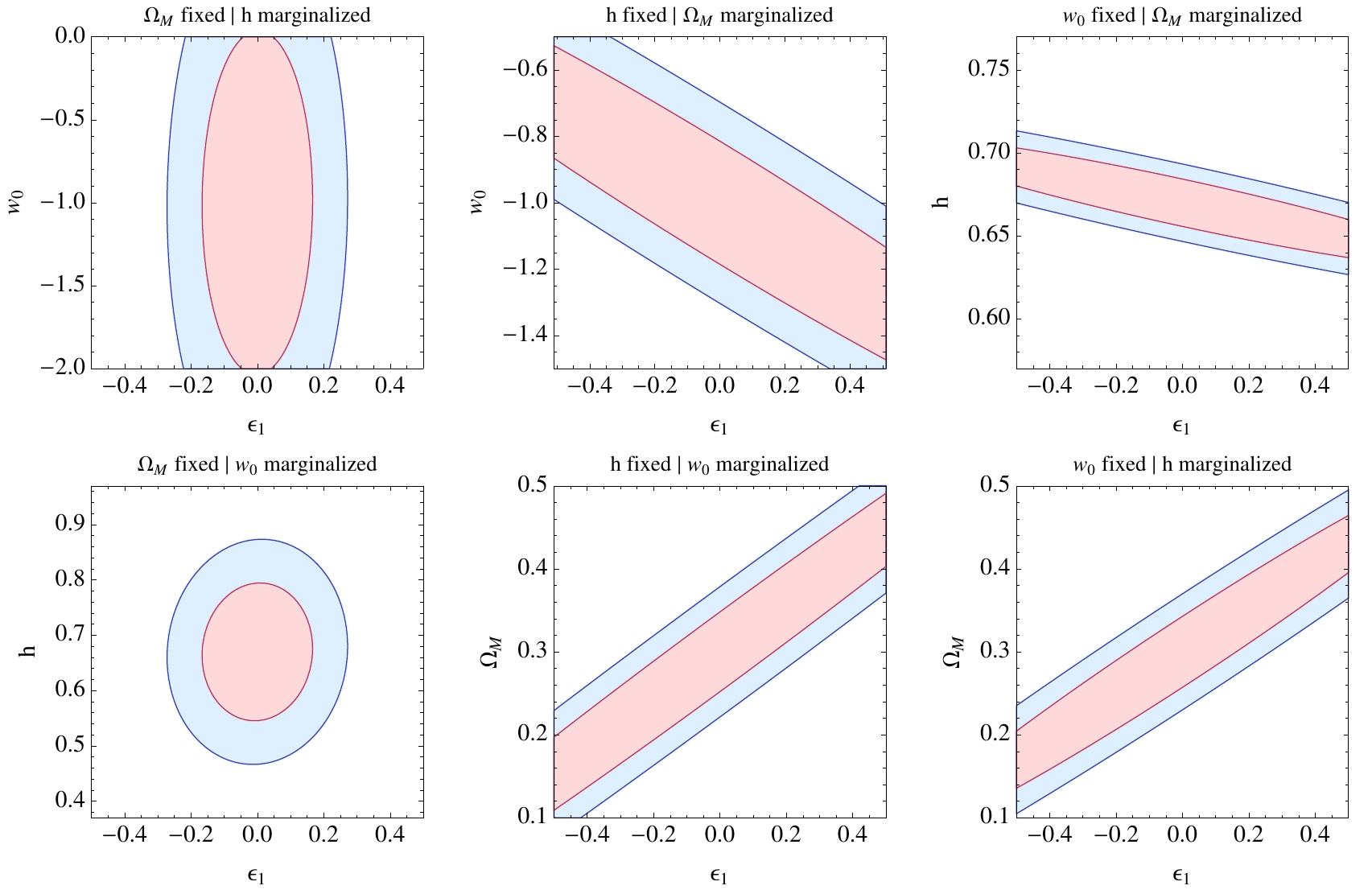}
\end{center}
\caption{Marginalized 1 (red) and 2$\sigma$ (blue) contour plots for IDE1 with three-parameter cosmological models and $z_i \rightarrow \infty$. Only results for the popIII BH models are shown.}
\label{fig:degeneracies_1}
\end{figure}
\begin{figure}
\begin{center}
	\includegraphics[width=\textwidth]{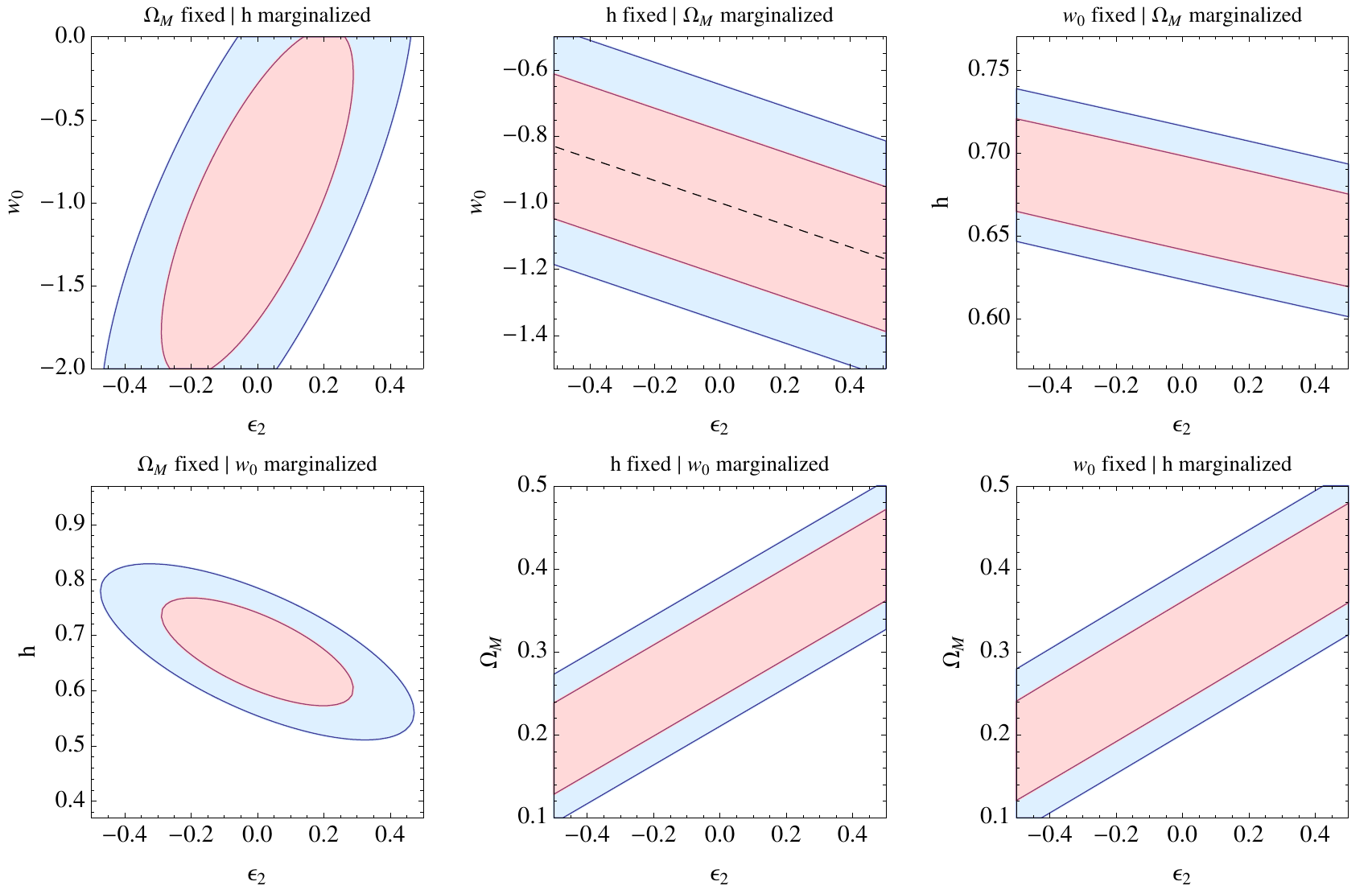}
\end{center}
\caption{Marginalized contour plots for IDE2 with three-parameter cosmological models and $z_i \rightarrow \infty$. Only results for the popIII BH models are shown. In the top-centre panel the black dashed lines denote the effective DE equation of state $w_0 + \epsilon_2/3 = -1$.}
\label{fig:degeneracies_2}
\end{figure}

The aim of this appendix is to visualise better the degeneracy between $\epsilon_1$ (or $\epsilon_2$) and $\Omega_{m}^0$ that occurs both in IDE1 and in IDE2 for the three parameter models. Fig.~\ref{fig:degeneracies_1} and Fig.~\ref{fig:degeneracies_2} show marginalised contour plots for the three three-parameter models where one parameter among $\Omega_{m}^0$, $h$ and $w_0$ has been fixed to its fiducial value and another one has been marginalised over. We set $z_i\rightarrow \infty$ since, as discussed in the main text, the degeneracy is less problematic at low $z_i$ thank to the properties of the distribution in redshift of the standard sirens outlined in the previous appendix. 
From the figures it appears clearly that fixing $\Omega_{m}^0$ breaks the degeneracy, while this does not happen in the other three-parameter models where $h$ or $w_0$ have been fixed to their fiducial values. Note that for IDE2 in the marginalized $(w_0, \epsilon_2)$ case, the degeneracy is perfectly aligned with the line $w_0 + \epsilon_2/3 = -1$. This occurs because in this model low-redshift data on the expansion of the universe can only constrain the effective DE equation of state given by $w_0 + \epsilon_2/3 = -1$, as explained for example in \cite{Murgia:2016ccp}, and as can be appreciated by taking the limit of small $z$ in Eq.~\eqref{eq:H_IDE2}.


\bibliographystyle{unsrt}
\bibliography{paper}

\end{document}